\documentclass[fleqn,usenatbib]{mnras}

\usepackage{newtxtext,newtxmath}
\usepackage{multicol}
\usepackage{enumitem} 

\usepackage[T1]{fontenc}

\DeclareRobustCommand{\VAN}[3]{#2}
\let\VANthebibliography\thebibliography
\def\thebibliography{\DeclareRobustCommand{\VAN}[3]{##3}\VANthebibliography}

\usepackage{graphicx}	
\usepackage{amsmath}	
\usepackage{subfig}     
\usepackage{placeins}   
\usepackage{float}      

\title[MHD simulations of protoplanetary discs I]{Global magnetohydrodynamic simulations of the inner regions of protoplanetary discs. I. Zero-net flux regime}

\author[M. J. O. Roberts, H. N. Latter and G. Lesur]{
Matthew J. O. Roberts$^{1}$\thanks{E-mail: mjor2@cam.ac.uk},
Henrik N. Latter$^{1}$
and Geoffroy Lesur$^{2}$
\\
$^{1}$Department of Applied Mathematics and Theoretical Physics, University of Cambridge, Centre for Mathematical Sciences, \\Wilberforce Road, Cambridge CB3 0WA, UK\\
$^{2}$Institut de Planétologie et d’Astrophysique de Grenoble, Université Grenoble Alpes, CNRS, IPAG, 38000 Grenoble, France
}

\date{Accepted XXX. Received YYY; in original form ZZZ}

\pubyear{\the\year}

\begin{document}
\label{firstpage}
\pagerange{\pageref{firstpage}--\pageref{lastpage}}
\maketitle

\begin{abstract}
The inner regions of protoplanetary discs, which encompass the putative habitable zone, are dynamically complex, featuring a well-ionised, turbulent active inner region and a poorly ionised `dead' outer region. In this first paper, we investigate a base-level model of the magnetohydrodynamic processes around the interface between these two regions, using five three-dimensional global magnetohydrodynamic simulations in the zero-net flux regime. We employ physically motivated profiles for Ohmic resistivity and ambipolar diffusion, alongside a simplified thermodynamic model comprising a cool disc and hot corona. Our results show that, first, large-scale coherent poloidal magnetic field loops form in the magnetorotational instability active region. These loops lead to the accumulation of tightly wound magnetic flux at the disc--corona temperature transition, driving strong, localised accretion flows in the surface layers of the active region. Second, an axisymmetric pressure maximum, extending across multiple disc scale heights, develops as a result of outward mass transport from the active region. This, in turn, triggers the Rossby wave instability and leads to the development of anticyclonic vortices. Third, the dead zone develops magnetic field with a distinct morphology, likely resulting from the outward diffusion of the large-scale poloidal loops in the active zone. This self-consistently generated field exhibits a vertical structure that can drive accretion in the inner dead zone via a weak magnetic-pressure wind. In the second paper in the series, we extend this work to the vertical-net flux regime, where global magnetic flux transport and magnetically driven outflows become dynamically significant.
\end{abstract}


\section{Introduction}

The inner regions ($\lesssim 2\,\text{au}$) of protoplanetary discs encompass several key transition zones -- including the star--disc boundary, silicate sublimation front, and dead--active zone interface -- that significantly impact the long and short time evolution of the disc. These regions are also critically important because they (a) include the habitable zone and are preferred sites for dust aggregation and terrestrial planet formation, (b) regulate protostellar assembly through (variable) disc accretion, and (c) contain the launching points of jets and magnetic outflows of different morphologies. \\
\indent The magnetohydrodynamic (MHD) processes operating within these regions are extremely complex. Beyond the critical radius of the dead--active zone interface ($\lesssim \!1\,\mathrm{au}$), turbulence driven by the magnetorotational instability \citep[MRI;][]{balbus_powerful_1991} is suppressed on account of the sharp drop-off in ionisation, leading to the formation of a quiescent MRI-dead zone \citep{gammie_layered_1996}. In this region, accretion is theorised to be driven by weak laminar magnetic-pressure winds \citep{bai_wind-driven_2013, gressel_global_2015}. The (possible) accretion mismatch between these regions drives the formation of large-scale disc structures -- rings and vortices -- that can trap dust and promote planetesimal growth \citep[e.g.][]{varniere_reviving_2006, kretke_assembling_2009}. At the same time, the transition between laminar and turbulent flow imposes a mismatch in magnetic flux transport \citep{guilet_global_2014, bai_hall_2017, leung_local_2019}. The magnetic flux influences disc evolution, by controlling the strength of MRI turbulence \citep{hawley_local_1995} and the launching and efficiency of magnetic outflows \citep[e.g.][]{lesur_systematic_2021}. In addition, the mass and magnetic flux advected through the active region must ultimately be loaded onto the protostar, and interact with its dipolar magnetic field \citep[e.g.][]{romanova_accretion_2015}, further contributing to the unsteady magnetic behaviour underlying observed accretion variability \citep{fischer_accretion_2023}. Finally, the interplay between the stellar jet, episodic turbulent magnetic outflows, and laminar magnetic outflows \citep[e.g.][]{tu_highly_2025}, likely influences the observed morphology of chemically nested conical outflows originating from the inner disc \citep[e.g.][]{de_valon_alma_2020, pascucci_nested_2024}. \\
\indent Direct observations of the inner regions of protoplanetary discs are essential to constrain this rich panoply of physics. However, they remain extremely challenging due to the proximity of the host star, the small angular resolution required, and the high opacity at wavelengths accessible to instruments like \textit{ALMA}. Nevertheless, interferometers, such as \textit{GRAVITY} and \textit{MATISSE} on the \textit{VLT}, have begun to provide critical insights on this region. For example, these instruments have revealed asymmetries in the inner disc of HD 163296 \citep[e.g.][]{varga_asymmetric_2021, gravity_collaboration_gravity_2021-1}, constrained the magnetospheric accretion process \citep{gravity_collaboration_gravity_2024-1}, and enabled the characterisation of hot dust near the silicate sublimation radius \citep{gravity_collaboration_gravity_2021}. However, they are unable to probe the inner disc midplane directly. \\
\indent Therefore, despite the inherent challenges of modelling protoplanetary discs \citep[see][]{haworth_grand_2016}, global numerical simulations of the inner disc \citep[e.g.][and references therein]{faure_thermodynamics_2014, flock_3d_2017, iwasaki_dynamics_2024} remain indispensable for studying disc evolution and interpreting recent observational advances. Encouragingly, the upcoming generation of high-resolution radio band observation instruments -- including the \textit{SKA} and the \textit{ngVLA} -- will, by the end of the decade, provide the sensitivity and resolution needed to directly probe the inner disc and support/verify theoretical models and simulations \citep[e.g.][]{ilee_observing_2020, ueda_probing_2022}. \\
\indent  This paper is the first in a series that investigates the inner regions of protoplanetary discs through GPU simulations performed with the \textsc{idefix} code \citep{lesur_idefix_2023}. We focus specifically on understanding the MHD processes around the dead--active zone interface using a simplified numerical setup: an initial zero-net flux (ZNF) magnetic field, physically motivated prescriptions for Ohmic and ambipolar diffusion, and a two-temperature structure representing a cool disc embedded in a hot corona. To maintain a controlled numerical setup, we omit several layers of physical complexity -- such as radiative transfer \citep[e.g.][]{gressel_global_2020}, dust, calculations of the ionisation fraction, and the Hall effect \citep[e.g.][]{bethune_global_2017} -- that may be important in global MHD simulations. The investigation of the ZNF regime in this paper will be followed by a study of the vertical-net flux regime in a subsequent paper, eventually progressing towards more comprehensive models incorporating dust, the Hall effect, and the interaction with the star. This staged approach will allow us to generalise our physical insights beyond the constrained parameter space accessible to global simulations. \\
\indent Finally, the paper is structured as follows. In Section~\ref{section:background} we provide the physical and numerical background that motivates and contextualises our simulations. Section~\ref{section:methods_and_model} outlines the inner disc model, key assumptions, and numerical methods. We then present the results of our simulations in three parts: the structure of the active zone and disc accretion in Section~\ref{section:accretion_and_disc_structure}; the formation and morphology of large-scale hydrodynamic (HD) structures at the dead--active zone interface in Section~\ref{section:large_scale_hydrodynamic_structures}; and the evolution of large-scale coherent magnetic fields in the dead zone in Section~\ref{section:magnetic_field_structures}. Concluding remarks are presented in Section \ref{section:conclusion}.


\section{Background and inner disc physics}
\label{section:background}

To establish the necessary context, we begin with a brief overview of the theory underlying the dead--active zone interface in Section~\ref{section:dead_active_zone_interface}. Section~\ref{section:temporal_variability} introduces the accretion paradigms and the role of the inner disc in driving accretion variability. This is followed by a discussion of the large-scale HD structures that emerge at the dead--active zone interface in Section~\ref{section:large_scale_structure_formation}. Finally, in Section~\ref{section:transport_magnetic_field}, we examine the origin and global transport of large-scale magnetic flux.

\subsection{The dead--active zone interface}
\label{section:dead_active_zone_interface}
Protoplanetary discs are weakly ionised systems, making them a unique archetype in the context of astrophysical discs \citep{lesur_magnetohydrodynamics_2021}. The seminal work of \cite{gammie_layered_1996} introduced the `layered accretion' paradigm, on account of Ohmic resistivity suppressing the MRI \citep[e.g.][]{jin_damping_1996}. This model posits that the MRI is active only in an `envelope', comprising a hot inner disc ($\lesssim 0.5$--$1\,\text{au}$) sufficiently ionised either thermally \citep[e.g.][]{umebayashi_chapter_1988} or thermionically \citep{desch_high-temperature_2015}, and disc surface layers, sufficiently ionised by stellar irradiation or cosmic rays. This envelope surrounds an MRI-dead interior. The complex interplay between the MRI, angular momentum transport, and localised accretion suggests that steady-state accretion onto the central star is unlikely within this framework \citep{gammie_layered_1996, terquem_new_2008}, implying that temporal variability is an intrinsic feature of the inner disc. \\
\indent However, a major limitation\footnote{The Hall effect, which we neglect in this work, also has impacts on dead-zone dynamics \citep[e.g.][]{lesur_thanatology_2014, bethune_global_2017, wurster_we_2021}.}of this model, is that it fails to account for the increasing role of ambipolar diffusion in the low-density surface layers surrounding the dead zone, which further suppresses MRI activity \citep[e.g.][]{blaes_local_1994, kunz_ambipolar_2004}. This led to a recharacterisation of the accretion paradigm in the dead zone -- from one driven by MRI turbulence in the surface layers \citep[e.g.][]{fleming_local_2003, turner_dead_2008}, to one governed by weak laminar magnetic winds \citep[e.g.][]{bai_wind-driven_2013, gressel_global_2015} -- as discussed further in Section~\ref{section:temporal_variability}. \\
\indent Despite its importance, there exists a limited amount of numerical work investigating the dead--active zone interface (see the taxonomy in Section \ref{section:modelling_dead_active_zone_interface}). Its location, shape and stability are highly complex due to the interpenetration of dynamics and thermodynamics \citep[e.g.][]{wunsch_two-dimensional_2006, latter_dynamics_2012, faure_thermodynamics_2014,cecil_variability_2024}. For instance, in the MRI-active region, temperature depends on the local dissipation provided by the turbulence, whilst the turbulence depends on the temperature through the thermal ionisation. Further feedback loops also arise from dust and mass accumulation, through their influence on opacity and ionisation, all of which can drive explosive dynamics and substantial variability in the accretion rate.

\subsection{Accretion and variability}
\label{section:temporal_variability}
An efficient and common mechanism for angular momentum is required to explain observed accretion rates of $\sim\!10^{-7}$--$10^{-8}\;\text{M}_{\odot}\,$yr$^{-1}$ \citep[e.g.][]{miotello_setting_2023} The prevailing view (see discussion in \citealt{lesur_hydro-_2023}) is that accretion in the poorly ionised disc ($r\gtrsim1\,\text{au}$) is driven by the vertical extraction of angular momentum by weak laminar magnetic-pressure winds \citep[e.g.][]{bai_wind-driven_2013, gressel_global_2015, bai_magneto-thermal_2016}, which remain effective even at very low field strengths, such as midplane poloidal plasma-$\beta$ values as high as $10^8$ \citep{lesur_systematic_2021}. In contrast, accretion in the better ionised region ($r\lesssim1\,\text{au}$) is facilitated by a combination of radial turbulent stresses from MRI turbulence \citep[e.g.][]{hawley_local_1995}, and radial laminar stresses from large-scale coherent magnetic fields \citep[e.g.][]{zhu_global_2018, jacquemin-ide_magnetic_2021}. However, this paradigm critically depends on the presence of a weak poloidal magnetic field threading the disc (see Section~\ref{section:transport_magnetic_field}). \\
\indent Protoplanetary discs also exhibit significant accretion variability in the form of strong outbursts that last for hundreds of years (e.g. FU Ori; \citealt{hartmann_fu_1996}). These major events obliterate the magnetospheric cavity and are often attributed to the dynamics of the dead--active zone interface, specifically, the activation of gravitational instability in the dead zone \citep{zhu_long-term_2010, bourdarot_fu_2023}, allied with a reactivation of the MRI, or possibly thermal instability \citep{bell_using_1994,cecil_variability_2024}. \\
\indent Recently, a broad spectrum of ubiquitous, smaller-amplitude variability has been identified \citep[see][]{fischer_accretion_2023}, with timescales linked to dynamical and viscous processes in the inner disc. This variability is classified by \citet{fischer_accretion_2023} as: (a) routine variability ($\lesssim 1$--$2$~mag), occurring on timescales of hours to days in the near-IR (magnetospheric) and up to years in the mid-IR (innermost disc); (b) bursts ($\sim 1$--$2.5$~mag), which are distinct, discrete and last from days to a year, likely linked to stochastic instabilities associated with the star--disc connection \citep{stauffer_csi_2016}; (c) episodic inner-disc processes; and (d) large-scale outbursts, like FU Ori and EX Lup, ($\sim 2.5$--$6$~mag), which develop over months to years and persist for decades. Recent observations of two outbursting systems, \textit{Gaia}~17bpi \citep{hillenbrand_gaia_2018} and \textit{Gaia}~18dvy \citep{szegedi-elek_gaia_2020}, provide particularly compelling evidence for the involvement of the inner disc \citep{cleaver_magnetically-activated_2023}. In both cases, the outbursts began in the near-IR before appearing in the optical, indicating an outside-in propagation through the inner disc. \\
\indent The broad range of accretion variability suggests that protoplanetary discs routinely undergo dynamical disruptions, further challenging the assumption of steady-state evolution \citep[e.g.][]{terquem_new_2008, mohanty_inside-out_2018, jankovic_mri-active_2021, jankovic_mri-active_2021-1, delage_steady-state_2022}. This underscores the need for global non-ideal MHD simulations of the inner disc, with careful consideration of the relevant timescales.
\subsection{Large-scale hydrodynamic structure formation}
\label{section:large_scale_structure_formation}
\indent In the absence of magnetic disc winds, the accretion mismatch across the dead--active zone interface inevitably leads to the formation of a localised pressure bump \citep[e.g.][]{dzyurkevich_trapping_2010}. This pressure maximum coincides with a vortensity extremum, usually sufficient to trigger the Rossby wave instability (RWI) \citep{lovelace_rossby_1999, lovelace_rossby_2014}, which saturates in coherent anticyclonic vortices. This behaviour has been confirmed by global simulations of the dead--active zone interface, including two-dimensional HD models \citep{lyra_planet_2009}, three-dimensional unstratified MHD models \citep{lyra_rossby_2012, faure_thermodynamics_2014}, and more advanced three-dimensional stratified MHD models incorporating realistic radiation physics \citep{flock_3d_2017}. \\
\indent These coherent structures make the dead--active zone interface a prime location for planetesimal formation, provided their formation mechanisms are robust enough to withstand the spectrum of variability intrinsic in the inner disc. For example, the accumulation of dust at the pressure maximum \citep[e.g.][]{kretke_assembling_2009, ueda_analytic_2017} has been proposed as a mechanism for forming multiple inner-region planets \citep{chatterjee_inside-out_2014, ueda_early_2021}. Anticyclonic vortices offer an additional pathway because they can trap dust in their cores \citep{barge_did_1995, tanga_forming_1996}, a process extensively studied in the context of Rossby-wave-induced vortices \citep[e.g.][]{meheut_dust-trapping_2012, miranda_long-lived_2017}. However, this mechanism is only effective if the vortices do not develop internal meridional flows \citep{meheut_rossby_2010} and are sufficiently long-lived -- avoiding cycles of formation and destruction \citep{faure_vortex_2015, flock_gaps_2015} or the onset of the elliptical instability \citep{pierrehumbert_universal_1986, lesur_stability_2009}, which can induce core turbulence or destroy the vortices.

\subsection{Provenance and transport of magnetic fields}
\label{section:transport_magnetic_field}
The presence and morphology of magnetic field structures are linked to the evolution and accretion dynamics of protoplanetary discs. Yet, due to current observational limitations, a complete picture for the magnetic field morphology remains elusive \citep{tsukamoto_role_2023}. Existing measurements tentatively suggest a weak vertical field dominated by a toroidal component, consistent with the configuration expected under the weak magnetic-pressure wind accretion paradigm in the dead zone. For instance, the circular polarisation of Zeeman-split CN molecular lines places tentative upper limits on the vertical magnetic field strength of $B_z < 0.8$~mG in TW~Hya \citep{vlemmings_stringent_2019} and $B_z < 6.1$~mG in AS~209 \citep{harrison_alma_2021}. Meanwhile, polarised thermal emission from aligned non-spherical dust grains implies a total magnetic field strength of $B\!\sim\!0.3$~mG in HD~142527, with the toroidal component dominating by a factor of four \citep{ohashi_observationally_2025}, although non-magnetic grain alignment mechanisms \citep[see][]{harrison_protoplanetary_2024} endanger this interpretation. \\ 
\indent The leading theory for the origin of a poloidal magnetic field threading the disc is inheritance from the protostellar nebulae \citep[e.g.][]{kunz_nonisothermal_2010,crutcher_magnetic_2012}, where magnetic flux is drawn inward during core collapse, shaping the field into a pinched, hourglass configuration. However, recent simulations of MRI-turbulent accretion discs by \citet{jacquemin-ide_magnetorotational_2024}, suggest that large-scale poloidal fields can also be generated \textit{in situ} via the MRI dynamo.\\
\indent Nevertheless, it is undoubtedly the case that the transport of such a large-scale magnetic field impacts disc evolution. The conventional advection-diffusion paradigm sets this magnetic flux transport as a competition between inward advection, due to bulk accretion flow, and outward diffusion, driven by turbulent or physical resistivity \citep[e.g.][]{lubow_magnetic_1994, heyvaerts_magnetic_1996}. Whilst extensive (semi-)analytic work has highlighted the importance of vertical disc structure \citep{rothstein_advection_2008, guilet_transport_2012, guilet_transport_2013} and non-ideal MHD effects \citep[e.g.][]{leung_local_2019}, a complete local model of flux transport remains elusive. As such, global numerical simulations are essential. \\
\indent Global simulations suggest an intriguing paradigm for magnetic flux transport in the inner regions of protoplanetary discs. In the ideal-MHD regime, recent studies show that advection dominates and flux is transported inward \citep{zhu_global_2018, jacquemin-ide_magnetic_2021}. Meanwhile, in the laminar, non-ideal-MHD regime, diffusion tends to dominate and typically drives flux outward \citep[e.g][]{bai_global_2017, gressel_global_2020, lesur_systematic_2021}, though this picture is complicated by the Hall effect \citep{bai_hall_2017}. Therefore, the transport of flux transport is divergent around the dead--active zone interface. Moreover, the fate of the inner magnetic field as it drifts inward and interacts with the stellar field remains an open question. As such, the global evolution of magnetic flux is expected to be strongly dependent on the interactions around the dead--active zone interface, with the star--disc boundary \citep[e.g.][]{pantolmos_magnetic_2020, takasao_connecting_2025, zhu_global_2025}, and the initial magnetic field configuration.

\section{Model and Numerical Methods}
\label{section:methods_and_model}
The global three-dimensional non-ideal MHD ZNF simulations in this study evolve the governing equations outlined in Section \ref{section:equations} as applied to the inner disc model detailed in Section \ref{section:inner_disc_model}. The corresponding numerical methods are presented in Section \ref{section:numerical_methods1}, followed by the key diagnostics in Section~\ref{section:diagnostics_and_definitions}.
\subsection{Governing equations}
\label{section:equations}
The evolution of our disc model is governed by the non-ideal MHD equations, in which Ohmic and ambipolar diffusion are included but the Hall effect neglected. These determine the evolution of the density $\rho$, velocity field $\mathbf{u}$, magnetic field $\mathbf{B}$, and energy density $E=e+\rho u^2/2+B^2/8\pi$, where $e$ is the internal energy density, through,
\begin{align}
    \partial_t{\rho} &= -\mathbf{\nabla} \cdot \left(\rho \mathbf{u}\right) 
     \label{equation:continuity_equation} 
     \\
    \partial_t{(\rho\mathbf{u})} &= - \mathbf{\nabla} \cdot \left( \rho\mathbf{u}\mathbf{u}\right)-\rho\mathbf{\nabla}\Phi - \mathbf{\nabla}{P} + \frac{\mathbf{J}\times\mathbf{B}}{c}
    \label{equation:momentum_equation} \\
    \partial_t\mathbf{B} &= -\nabla \times \left(\mathbf{\mathcal{E}_I} + \mathbf{\mathcal{E}_{NI}}\right) \label{equation:induction_equation} \\
    \label{equation:energy_equation}
    \partial_tE &= - \nabla \cdot \left[ \left(E + P +\frac{B^2}{8\pi}\right) \mathbf{u} - \frac{\mathbf{B} (\mathbf{B} \cdot \mathbf{u})}{4\pi} -\frac{\mathbf{\mathcal{E}_{NI}}\times\mathbf{B}}{4\pi}\right]\\&\;\,\,\;-\rho \mathbf{u} \cdot \nabla \Phi + \mathcal{L} \nonumber,
\end{align}
where the ideal and non-ideal components of the electromotive field are given by, respectively, 
\begin{equation}
	\mathbf{\mathcal{E}_{I}} = \mathbf{u}\times\mathbf{B} \;\; \textrm{and} \;\; \mathbf{\mathcal{E}_{NI}} = \frac{4\pi}{c} \left[  - \eta_\text{O} \mathbf{J}+ \eta_\text{A} (\mathbf{J} \times \mathbf{b}) \times \mathbf{b} \right],
\end{equation}
under the Gaussian unit convention. Here, $\Phi=-GM/r$ is the gravitational potential of a central spherical star of mass $M$ in spherical coordinates  $(r,\theta,\phi)$, $P$ the gas pressure, $c$ the speed of light, $\mathbf{J}= (c/4\pi)\mathbf{\nabla}\times\mathbf{B}$ the current density, $\mathcal{L}$ an imposed energy-density thermal-relaxation term, $\mathbf{b}$ the direction of the magnetic field, $\eta_{O}$ the Ohmic diffusivity, and $\eta_\text{A}$ the ambipolar diffusivity. \\
\indent The system of equations is closed using the equation of state for a perfect gas, $P=e(\gamma-1)$, where $\gamma$ is the adiabatic index. To ensure that the integration scheme is positive-definite with respect to $P$, particularly in the low-density regions, we adopt $\gamma=1.05$. The disc temperature is controlled through a $\beta$-cooling prescription: the system is relaxed towards a target temperature, $T_\textrm{eff}(R,z)$, using,
\begin{equation}
    \mathcal{L}= \frac{P - {T_{\textrm{eff}}\rho}}{\tau(\gamma-1)}.
    \label{equation:temperature_beta_model}
\end{equation}
 The characteristic thermal-relaxation timescale is $\tau=0.1\Omega_\text{K}^{-1}$, so that the vertical shear instability is marginal \citep{nelson_linear_2013}, where $\Omega_\text{K}^2=GMR^{-3}$ is the squared Keplerian angular frequency. This rapid thermal readjustment leads to an effectively quasi-local isothermal equation of state, $P=c_\text{s}^2\rho$, where the local isothermal sound speed in the disc is $c_\text{s}=H\Omega_\text{K}$, for the disc scale height $H$. \\
 \indent Finally, cylindrical coordinates $(R,\phi,z)$ are also used in this work, motivated by the axisymmetric nature of the imposed dead--active zone interface.

\subsection{Inner-disc model}
\label{section:inner_disc_model}
Our model of the inner disc regions is characterised by the following features: (i) it is magnetised with a dead--active zone interface at a fixed cylindrical radius, modelled using prescribed profiles of Ohmic diffusion and ambipolar diffusion; (ii) it is cool and embedded in a hot, low-density corona; (iii) apart from the central gravitational potential, the disc is dynamically isolated from the star; (iv) it neglects both self-gravity and explicit viscosity; and (v) it is dust-free.
\begin{figure*}
    \centering  
    \subfloat{\includegraphics[height=0.33\textwidth]{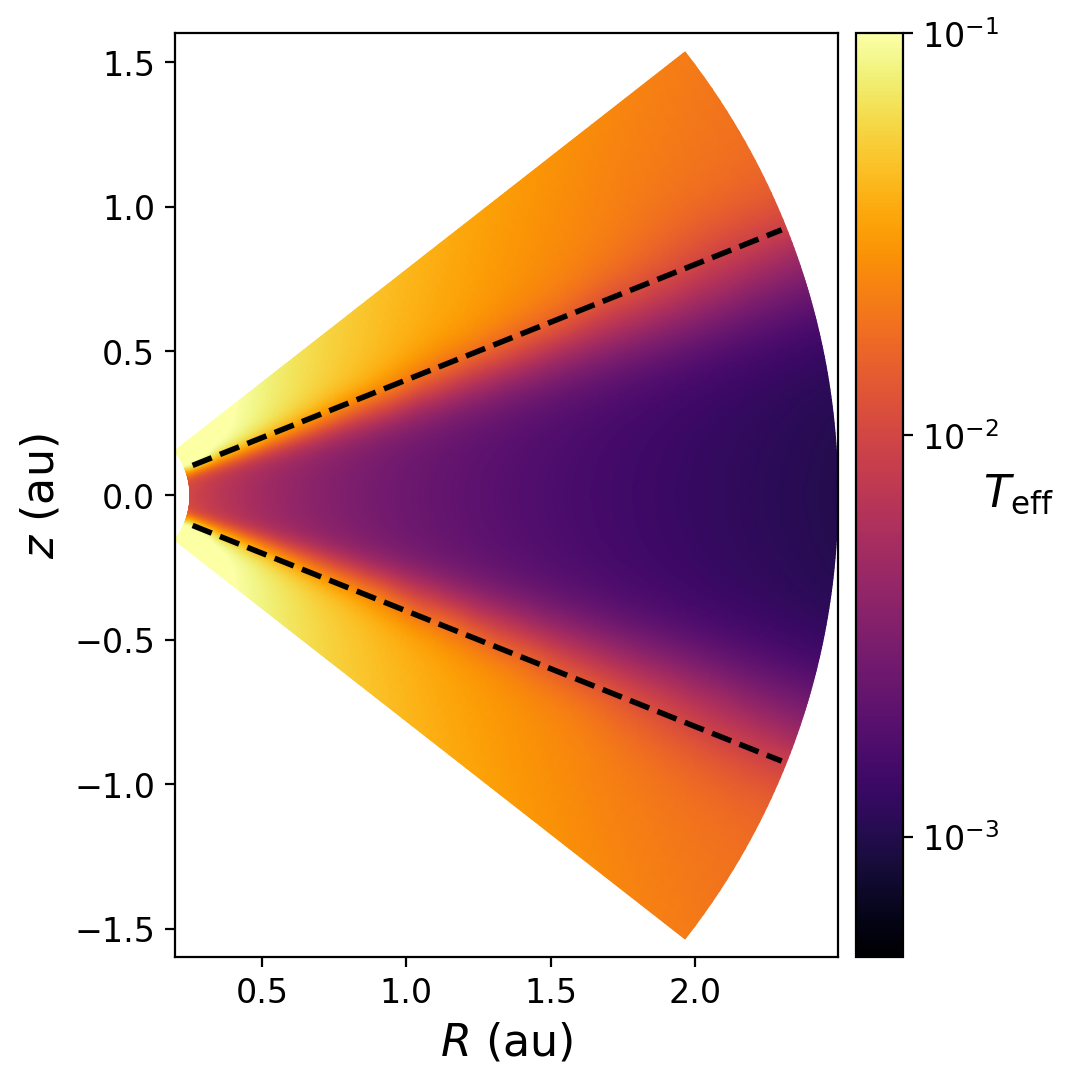}} 
    \subfloat{\includegraphics[height=0.33\textwidth]{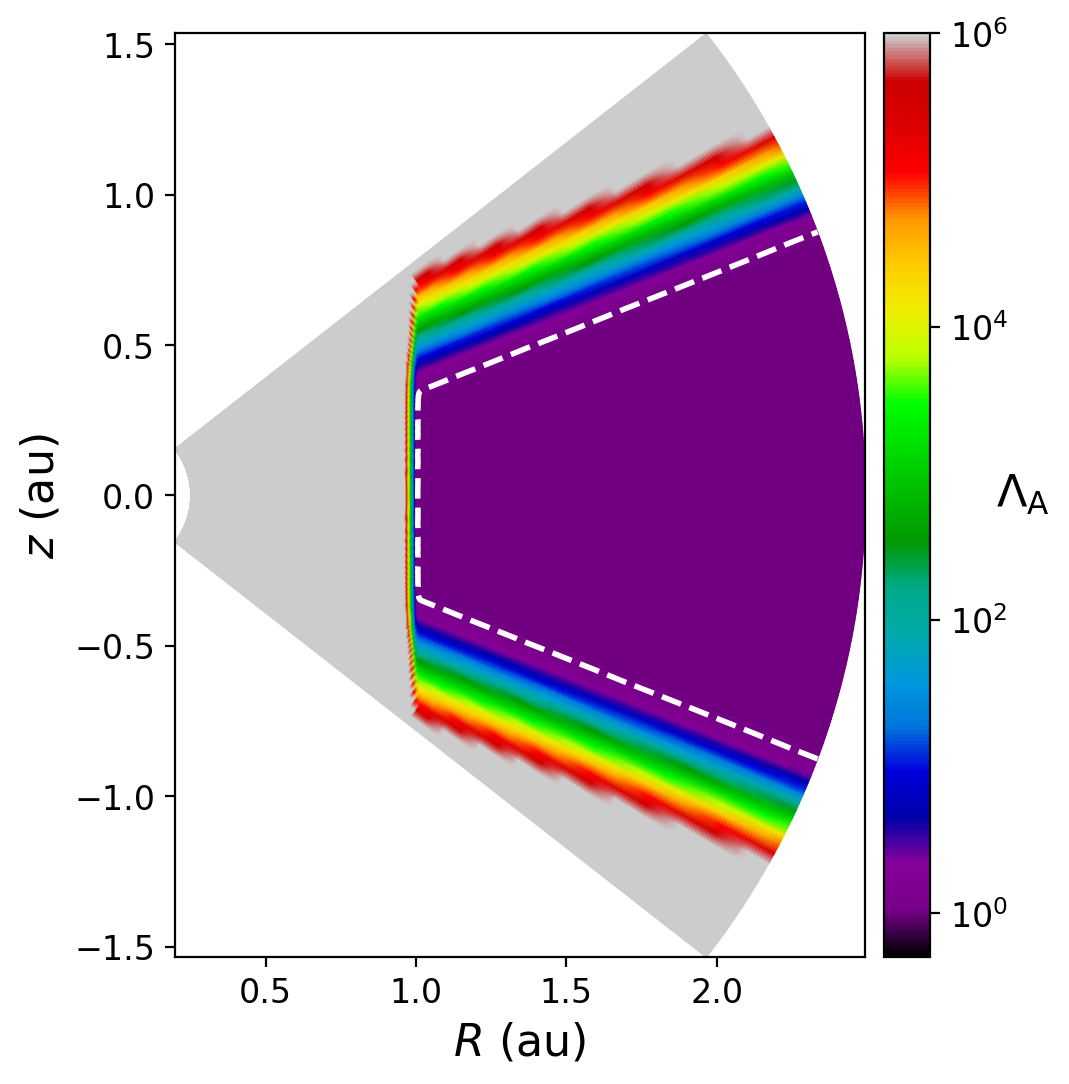}} 
    \subfloat{\includegraphics[height=0.33\textwidth]{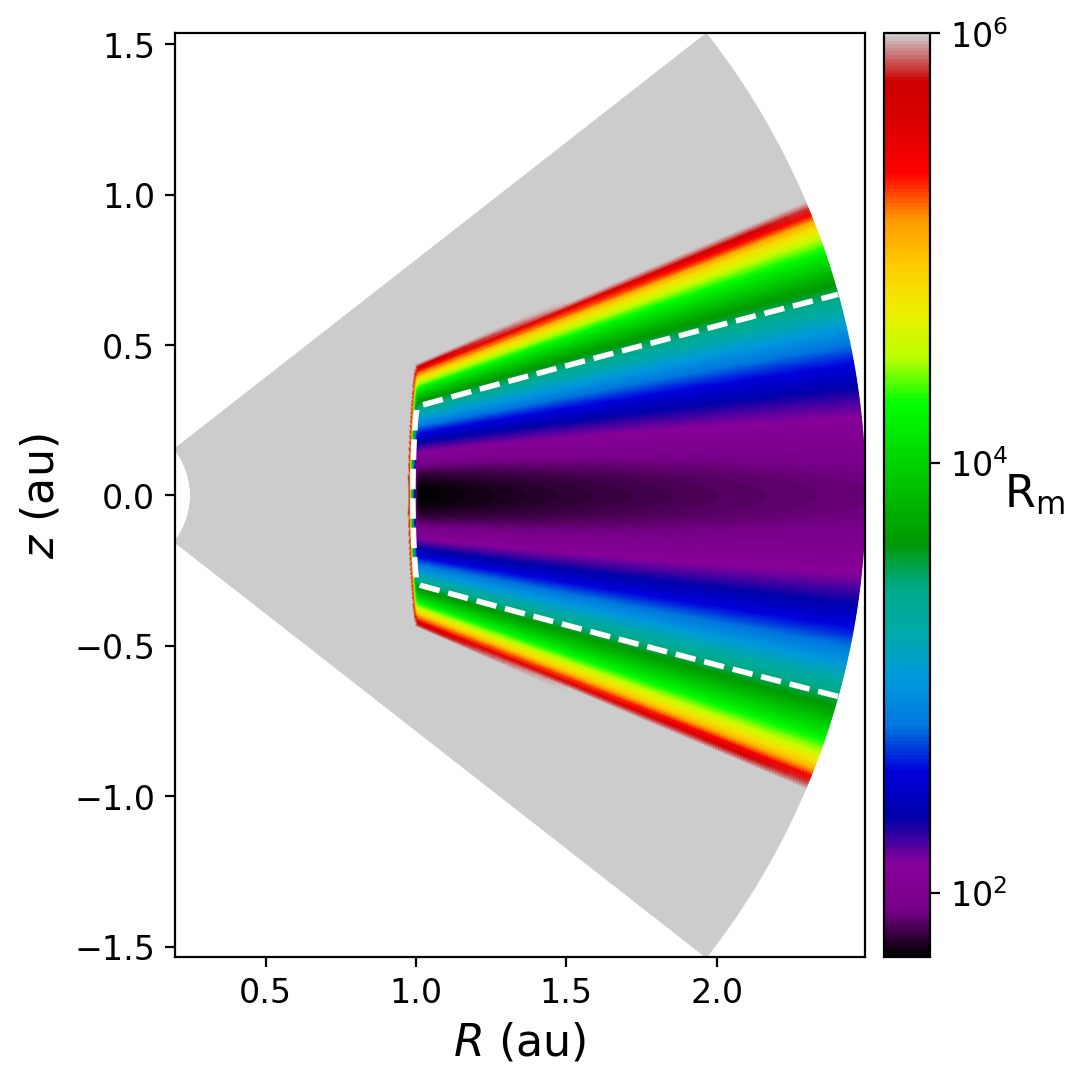}} 
    \caption{Meridional ($R,z$) plots of the prescribed target temperature $T_{\text{eff}}$ \eqref{equation:target_temperature} (left), ambipolar Elsässer number $\Lambda_\text{A}$ (middle), and initial magnetic Reynolds number $\text{R}_\text{m}$ (right), as described in Section \ref{section:inner_disc_model}. The dashed black lines in the left panel mark the disc--corona transition at $z=\pm 4H$. The dashed white lines show empirical thresholds for stratified ZNF MRI stability: R$_m\lesssim 3000$ \citep{flock_turbulence_2012} and $\Lambda_\text{A}\lesssim 1$ \citep{bai_effect_2011, simon_turbulence_2013}, which is less certain. These clearly delineate the dead--active zone interface at $R_{\text{DZI}}=4R_0=1\,\text{au}$. For clarity, the non-ideal MHD diffusivities are limited to $10^6$, and the Ohmic-resistive inner radial buffer region is not shown.}
    \label{fig:non_ideal_MHD_setup}
\end{figure*}
\subsubsection{Temperature prescription}
\label{section:thermodynamics}
The target temperature profile for the $\beta$-cooling prescription \eqref{equation:temperature_beta_model},
\small
\begin{equation}
    T_{\textrm{eff}}(R,z) = \frac{1}{2}\left[ \left(T_{\textrm{disc}}+T_{\textrm{cor}}\right)+\left(T_{\textrm{disc}}-T_\textrm{cor}\right)\tanh\left(\frac{\left|z\right| - 4\varepsilon R_\text{i}}{R_0} \right)\right],
    \label{equation:target_temperature}
\end{equation}
\normalsize
\noindent
is exhibited in Fig. \ref{fig:non_ideal_MHD_setup}. The model splits the domain into a cold dense disc with temperature $T_\textrm{disc}(R) = T_0R_0\varepsilon^2/R_\text{i}$, and a hot, low-density corona with temperature $T_\textrm{cor}(R) = 16T_\textrm{disc}(R)$. Here $\varepsilon=H/R$ is the disc aspect ratio, the subscript $0$ refers to the midplane at the inner edge of the domain, and $R_\text{i}=\max(R,R_0)$, which ensures consistency with future models that are closed at the poles. Thus the radial profile of the effective temperature scales as $T\propto R^{-1}$ when $R\geq R_0$. This enforces a constant aspect ratio. We adopt $\varepsilon=0.1$ as a compromise between numerical efficiency -- ensuring that the MRI is adequately resolved -- and the expected inner disc thickness of roughly 0.03--0.05 at $R=1$ au \citep[e.g.][]{bitsch_structure_2015}.\\
\indent This simplified temperature profile prescribes a hot corona to account for the mixture of stellar irradiation \citep[e.g.][]{aresu_x-ray_2011} and the inefficient gas cooling resulting from reduced thermal accommodation on dust particles in this dust-sparse region \citep[see fig. 10 in][]{thi_radiation_2019}. The transition height is fixed at $z=\pm4H$, following the ionisation model presented in fig. 1 of \cite{lesur_thanatology_2014}.
\subsubsection{Non-ideal MHD model and the dead--active zone interface}
\label{section:modelling_dead_active_zone_interface}
To capture the complex dynamic and thermodynamic behaviour of the dead--active zone interface, a broad range of numerical models have been used in previous work. These include: (i) spatially static HD viscosity transitions \citep[e.g.][]{lyra_planet_2009, miranda_rossby_2016, ueda_dust_2019}; (ii) spatially static Ohmic resistivity transitions \citep[e.g.][]{dzyurkevich_trapping_2010,okuzumi_modeling_2011, lyra_rossby_2012}, sometimes extended to include ambipolar diffusion \citep{iwasaki_dynamics_2024}; and (iii) temperature-dependent transitions in either viscosity \citep[e.g.][]{ueda_dust_2019, cecil_variability_2024} or Ohmic resistivity \citep[e.g.][]{faure_thermodynamics_2014, faure_vortex_2015, faure_planet_2016, flock_3d_2017}. \\
\indent Consistent with the second class of models described above, we adopt a spatially static, physically motivated non-ideal MHD setup that includes both Ohmic and ambipolar diffusion to model the dead--active zone interface. The magnetic Reynolds and ambipolar Elsässer numbers are defined as,
\begin{equation}
\text{R}_\text{m} = \frac{\Omega_\text{K} H^2}{\eta_\text{O}}
\quad \text{and} \quad
\Lambda_\text{A} = \frac{v^2_\text{A}}{\Omega_\text{K} \eta_\text{A}},
\label{equation:non_ideal_MHD_diffusivities}
\end{equation}
where $v_\text{A}=B/\sqrt{4\pi\rho}$ is the Alfvén speed. Their profiles are prescribed as,
\small
\begin{align}
 \label{equation:non_ideal_MHD_prescription}
 \text{R}_\text{m}(R,z) =& \,\frac{4\text{R}_{\text{m}_0} \rho_0 R_0}{\rho R} 
\left[1 - \tanh\left(\frac{|z| - 4\varepsilon R}{0.2 R_0}\right)\right]^{-1} \\ &\times
\left[1 + \tanh\left(\frac{R - R_{\textrm{DZI}}}{0.02 R_0}\right)\right]^{-1},  \nonumber \\
\Lambda_\text{A}(R,z) =& \,4\Lambda_{\text{A}_0}\left[1 - \tanh\left(\frac{|z| - 4\varepsilon R}{0.2 R_0}\right)\right]^{-1} \left[1 + \tanh\left(\frac{R - R_{\textrm{DZI}}}{0.02 R_0}\right)\right]^{-1} \nonumber, 
\end{align}
\normalsize
\noindent
as shown in Fig.~\ref{fig:non_ideal_MHD_setup}, with $\Lambda_{\text{A}_0} = 1$ and $\text{R}_{\text{m}_0} = 25$ guided by estimates from fig. 8 in \citet{thi_radiation_2019} and fig. 10 in \citet{lesur_magnetohydrodynamics_2021}. We take the dead--active zone interface to be at ${R_{\textrm{DZI}}}=4R_0$. The $\text{R}_\text{m}$ scaling with $\rho$ and $R$ follows \citet{lesur_systematic_2021}, assuming constant $\Lambda_\text{A}$, a two-species plasma (electrons and ions), and dominant ions, primarily \textsc{HCO}$^+$, that are much heavier than the main neutral component, \textsc{H}$_2$. \\
\indent Therefore, the model assumes that ionisation in the surface layers becomes effective at the same height as the disc--corona temperature transition, $|z|\gtrsim4H$, and it imposes a static cylindrical dead-zone interface, $R=R_\text{DZI}$. This interface is sharply defined to reflect the steep radial variation in the ionisation fraction in the inner disc, whether driven by thermal ionisation of alkali metals (\textsc{N}a, \textsc{K}, etc.) via the Saha equation or by thermionic and ion emission from dust grains \citep{desch_high-temperature_2015, williams_ionization_2024}. 
\subsection{Numerical methods}
\label{section:numerical_methods1}
The global non-ideal MHD simulations presented in this work are performed using the \textsc{idefix} code, which integrates the compressible MHD equations using a finite-volume high-order Godunov method. The interpolation to cell interfaces is performed using a linear piecewise reconstruction scheme with a van Leer slope limiter, and inter cell-fluxes are computed using the HLLD approximate Riemann solver \citep[for motivation see][]{flock_turbulence_2011}. The quantities are evolved in time using a total-variation diminishing second order Runge--Kutta method with the Courant number set to $\text{C}=0.5$. The solenoidal condition ($\mathbf{\nabla}\cdot\mathbf{B}=0$) is enforced to machine precision via the constrained transport method \citep{evans_simulation_1988} and an electromotive-field reconstruction based on the $\mathcal{E}^\text{c}$ scheme \citep{gardiner_unsplit_2005}. Parabolic non-ideal MHD diffusivity terms are handled by a centred finite-difference method. All simulations were run on NVIDIA A100 GPUs using double-precision arithmetic.\footnote{\texttt{ZNF-FID} used roughly 800 GPU-hours to perform $6.2\times 10^{6}$ integration cycles, achieving $1.8\times 10^{9}$ cell updates in total per second across 16 GPUs, using a 4–1–4 decomposition in $(r,\theta,\phi)$.} \\
\begin{table}
\captionof{table}{Numerical properties of the five three-dimensional ZNF global MHD simulations. Each is integrated up to $t_{\textrm{in}}^{\textrm{end}}$ on a domain with azimuthal extent $\phi_{\textrm{max}}$. \texttt{ZNF-MRI-IC} refers to a simulation where the non-ideal MHD profile is imposed on top of a fully MRI-turbulent disc at $t_{\textrm{in}}=0$, whilst \texttt{ZNF-AZ} refers to a simulation where the disc is in the ideal-MHD regime, such that there is no exterior dead zone, or dead--active zone interface. The total computational cost of these five production simulations was approximately $3000$ GPU hours.}
\centering
\begin{tabular}{|c c c c c|} 
 \hline
 Model name & $\left(N_r,N_\theta,N_\phi\right)$ & $\phi_{\textrm{max}}$ & $t_{\text{in}}^{\text{end}}$ & Reason \\ [0.5ex] 
 \hline
\texttt{ZNF-FID} & $(512,192,512)$ & $2\pi$ & 750 & fiducial \\
\texttt{ZNF-HRES} & $(1024,192,1024)$ & $2\pi$ & 375 & resolution\\
\texttt{ZNF-LONG} & $(512, 192,256)$ & $\pi$ & 1800 & late time\\
\texttt{ZNF-MRI-IC} & $(512,192,256)$ & $\pi$ & 750 & initial condition  \\
\hline
\texttt{ZNF-AZ} & $(512,192,256)$ & $\pi$ & 750 & full active \\
\hline
\end{tabular}
\label{table:list_global_simulations}
\end{table}
\indent In total, we present five production simulations: a fiducial simulation, \texttt{ZNF-FID}, and four extensions, each designed to test the robustness of the results by varying the numerical domain or the initial conditions (see Table~\ref{table:list_global_simulations} for grid details). \texttt{ZNF-HRES} serves as a resolution test. \texttt{ZNF-LONG} explores longer-term evolution. \texttt{ZNF-AZ} removes the non-ideal MHD diffusivity prescription, such that the disc is fully MRI turbulent. \texttt{ZNF-MRI-IC}, begins from a fully MRI-turbulent initial condition, onto which the non-ideal MHD diffusivity profile is imposed at $t_{\text{in}}=0$.

\vspace{1mm}
\FloatBarrier
\noindent
\subsubsection{Initial conditions}
\label{section:initial_conditions}
The initial HD fields are set to obey approximate radial and vertical hydrostatic equilibrium in the disc (after neglecting the Lorentz force) via \citep[e.g.][]{nelson_linear_2013},
\begin{align}
    \rho(R,z) &= \rho_0\left(\frac{R_\text{i}}{R_0}\right)^{-3/2}\exp\left[\frac{GM}{c_\text{s}^2}\left(\frac{1}{r}-\frac{1}{R}\right)\right], 
    \label{eq:density_initial_condition} \\
    \label{eq:uphi_initial_condition}
    u_\phi(R,z)  &=
    \begin{cases}
    \left(\frac{GM}{R}\right)^{1/2} \sqrt{\frac{R}{r} - \frac{5}{2}\left(\frac{H}{R}\right)^{\scriptscriptstyle 2} } , & \text{if $R\geq R_0$} \\
    \rule{0pt}{2.0em}
    R\left(\frac{GM}{R_{0}^3}\right)^{1/2}, & \text{otherwise},      \end{cases}
\end{align}
where the midplane gas density profile has been set in a self-similar fashion such that the initial gas surface density $\Sigma=\int^\infty_{-\infty}\rho\,\text{d}z\propto R^{-1/2}$, compatible with some observations \citep[see][]{williams_measuring_2016, miotello_setting_2023}. \\
\indent Note that the above expressions are not at equilibrium in the corona because of the larger temperature in this region and that the configuration has been chosen to provide consistency with future models that are closed at the poles. The remaining velocity components, $u_{r}$ and $u_\theta$, are initialised with slightly sub-sonic white noise in the inner disc. \\
\indent The initial ZNF magnetic field configuration is purely toroidal and confined to the active zone via
\begin{align}
B_\phi(R, z) = &\sqrt{\dfrac{2\rho_0 R_0^{3/2}c_\text{s}^2}{\beta_0 R^{3/2}}}\sin\left[ 4\pi \left( \dfrac{R}{R_0} - 1.5 \right) \right] \label{eq:magnetic_initial_condition} \\ &\times\max\left(1 - \dfrac{z^2}{4 H^2},\ 0 \right),\;\;\;\;\;\;\;\text{if}\; \;1.5R_0<R<4R_0 \nonumber,
\end{align}
and otherwise it is set to zero. The initial (minimum) midplane plasma-beta parameter is set to $\beta_0=20$. This configuration, inspired by previous global ZNF studies \citep[e.g.][]{flock_turbulence_2011, sorathia_global_2012, parkin_global_2013, parkin_global_2014}, ensures that saturated MRI turbulence can efficiently sustain a dynamo over the simulation timescale. We stress that initially there is no poloidal magnetic field, and that the dead zone is free of magnetic fields.
\subsubsection{Grid, resolution and units}
\label{section:grid_resolution_code_units}
\begin{figure*}
    \includegraphics[width=0.9\textwidth]{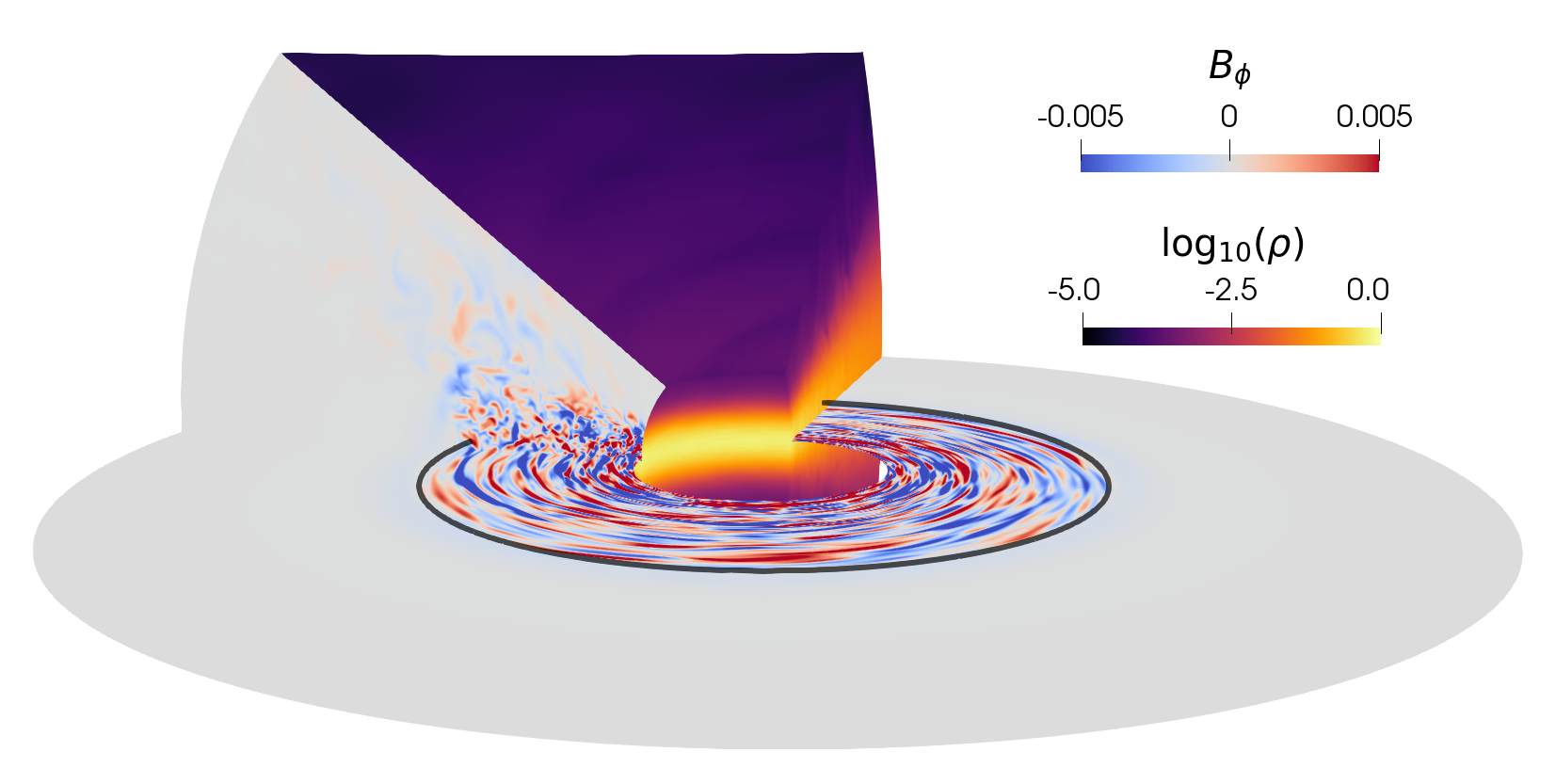}
    \caption{Snapshot of \texttt{ZNF-FID} at $t_{\textrm{in}}=30$, showing a slice through the disc midplane embedded within a quarter-azimuthal wedge of the full domain. The toroidal magnetic field, $B_\phi$, highlights the self-consistent turbulent magnetic structures, clearly delineating the midplane location of the dead--active zone interface (black line) at $R_{\text{DZI}}=4R_0 =1$ au. Over time, a complex laminar $B_\phi$ structure extends throughout the entire dead zone (see Fig.~\ref{fig:bx3_evolution} and Section~\ref{section:magnetic_field_structures}). The remaining faces of the wedge show the density $\rho$, revealing the vertically stratified disc embedded into a low-density corona.}
    \label{fig:comparison_3D_plot}
\end{figure*}

The numerical domain is described here for the fiducial simulation \texttt{ZNF-FID}, with changes made in the other four simulations outlined in Table \ref{table:list_global_simulations}. In spherical radius, a logarithmic grid with 512 cells spans $r\in[R_0,10R_0]$, such that the radial cell width scales as $\delta{r}\propto{r}$. The azimuthal domain covers $\phi \in [0, 2\pi]$ with 512 uniformly spaced cells. In the meridional direction, 128 uniformly spaced cells span $\theta \in [1.27, 1.87]$, corresponding to $z = \pm 4H$ about the midplane. This is extended by two geometrically stretched buffer regions of 32 cells each in $\theta \in [0.90, 1.27]$ and $\theta \in [1.87, 2.25]$, respectively. Therefore, the resulting computational domain is a spherical wedge extending to $z=\pm 8H$ in the meridional direction, as illustrated by the quarter-azimuthal wedge rendering shown in Fig. \ref{fig:comparison_3D_plot}. \\
\indent We define two reference timescales:  $t_{\text{in}}=2\pi/\Omega_\text{K}(R_0)$ corresponding to the orbital period at the innermost radius, and $t_{\textrm{DZI}}=2\pi/\Omega_\text{K}(R_{\textrm{DZI}})=8 t_{\text{in}}$, corresponding to the orbital period at the dead--active zone interface. Code units are chosen such that $GM=\rho_0=T_0=R_0=1$. \\
\indent Finally, the scale-free nature of the setup is broken by the introduction of the dead--active zone interface, which imposes a fixed physical location for the sharp ionisation transition in the inner disc. Consequently, we normalise the length scales in all the results by ${R_{\textrm{DZI}}}=4R_0 =1$ au, thereby consistently emphasising that this model applies only to the inner disc.\footnote{Whilst numerically convenient, the choice of $\varepsilon=0.1$ leads to an effective midplane temperature roughly 2--3 times higher than expected at $R=R_\text{DZI}$. However, this has no bearing on our results, given the relative simplicity of our disc model.}
However, we stress that $R_0=1 = 0.25\,\text{au}$ for the purposes of interpreting the analysis of the fields, which are presented in code units.
\subsubsection{Boundary and internal conditions}
At the inner radial boundary condition, $\rho$, $P$, $u_\theta$, $B_\theta$ and $B_\phi$ are copied from the innermost active cell into the ghost cells, whilst $u_\phi=R\Omega_{K}(R_0)$, thereby imposing solid-body rotation at the inner hole of the spherical domain. To prevent inflow from the boundary, $u_r$ is copied into the ghost cells if $u_r<0$ in the first active cell; otherwise, it is reflected symmetrically across the boundary. At the outer radial boundary and the meridional boundaries, the outflow conditions prescribed in \textsc{idefix} are applied, enforcing zero-gradients for all variables, except that when $u_r \geq 0$ in the first active cell, $u_r$ is set to zero. Finally, the azimuthal boundaries are periodic. \\
\indent These boundary conditions are completed by a buffer zone near the inner radius ($r<1.5R_0$) to mitigate boundary artefacts \citep{dzyurkevich_trapping_2010}. Within this region, $\mathbf{u}$ and $\rho$ are relaxed to their initial values on a timescale of 0.1 local orbital periods. In addition, an Ohmic-resistive damping term, $\eta_\text{O} = 0.05\varepsilon^2\max(1.5R_0 - r, 0)$, is applied to suppress magnetic fluctuations \cite[e.g.][]{fromang_global_2006}. Finally, to avoid restrictive time steps in the low-density corona, a density floor of $\rho_\text{min} = 10^{-9}\rho_0$ is imposed, and the Alfvén speed is limited to $v_\text{A} \leq 10 \,R_0\Omega_{K}(R_0)$, by adjusting the local density rather than the magnetic field.
\subsection{Diagnostics and definitions}
\label{section:diagnostics_and_definitions}
\subsubsection{Averages and fields}
The system around the dead--active zone interface evolves dynamically, exhibiting both laminar and turbulent behaviour across a range of time and length scales. To capture this complexity, we define the following integrals for a generic field $Q$:
\begin{align}
    \langle Q \rangle _\phi &= \frac{1}{\phi_{\textrm{max}}}
\int^{\phi_{\textrm{max}}}_{0} Q \;d\phi, \\
    \langle Q \rangle _{\phi,t} &=  \frac{1}{t_2-t_1} \int^{t_2}_{t_1} \langle Q \rangle_\phi\;dt, 
    \label{equation:spacetime_averages} \\
    \langle Q \rangle_{\phi,nH} &= \int^{\theta(z=-nH)}_{\theta(z=+nH)} r \sin(\theta)\langle Q\rangle_\phi\;d\theta,
\end{align}
where the temporal average is computed over the interval $[t_1,t_2]$ with a resolution of $\Delta t = 0.25 t_{\text{in}}$, and the meridional integration is taken over $n$ scale heights either side of the midplane. \\
\indent The poloidal component  $\mathbf{Q_p}$, for a three-dimensional field, is defined via the decomposition:
\begin{equation}
    \mathbf{Q} = \mathbf{Q_p} + Q_\phi \mathbf{e_\phi},
    \label{eqn:poloidal_field_definition}
\end{equation}
where $\mathbf{e_\phi}$ is the azimuthal basis vector. Line integral convolution (LIC) is used to aid visualisation of two-dimensional vector fields.

\subsubsection{Azimuthal asymmetries}
\label{section:averages_asymmetries_magnetic_fields}
\indent To diagnose the non-axisymmetric components, a Fourier decomposition of $Q$ into azimuthal ($m$) modes at the midplane is defined as,
\begin{equation}
    Q_m(R,t) = \frac{1}{\phi_{\textrm{max}}}\int^{\phi_{\textrm{max}}}_0 Q(r, \theta=\pi/2, \phi, t) e^{-i m \phi} d\phi,
    \label{eq:fourier_m_mode_decomposition}
\end{equation} 
with an associated power spectrum of $P_{Q,m}(R,t) =\left| Q_m(R,t) \right|^2.$ The vorticity of the velocity field is $\boldsymbol{\omega}=\nabla\times\mathbf{u}$ and the vortensity is,
\begin{equation}
    \xi = \frac{(\nabla\times\mathbf{u})_z}{\Sigma},
\label{equation:vortensity_definition}
\end{equation}
which is simply a ratio of the vertical component of the vorticity to the surface density. The deviation of the azimuthal velocity from Keplerian flow, $ u_\text{K} = R \Omega_\text{K} $, is defined as $u_\phi' = u_\phi - u_\text{K}$.
\subsubsection{Accretion and stresses}
\label{section:accretion_mechanism}
Within the turbulence and wind-driven accretion paradigms outlined in Section~\ref{section:temporal_variability}, stresses -- both turbulent and laminar -- are essential for facilitating angular momentum transport, enabling accretion through the disc \citep[see][]{lesur_magnetohydrodynamics_2021}.\\
\indent Throughout the rest of this section, $\langle \cdot \rangle$ denotes a generic average, and its specific meaning is clearly stated when used in the results. The total, averaged spherical radial Reynolds and Maxwell stresses \citep[e.g.][]{mishra_strongly_2020} are defined as,
\begin{align}
    \mathcal{R}_{r\phi} &= \langle\rho\delta{u_r}\delta{u_\phi}\rangle, \\
    \mathcal{M}_{r\phi} &= \frac{-\langle B_r B_\phi\rangle}{4\pi},
    \label{eq:maxwell_rphi_stress}
\end{align}
where $\delta Q$ denotes the fluctuation of $Q$ relative to its spatio-temporal average: $\delta Q = Q -\langle{Q}\rangle$. Following the seminal work of \citet{shakura_black_1973}, these can be written in dimensionless form by introducing an $\alpha$ parameter:
\begin{equation}
    \alpha = \frac{\mathcal{R}_{r\phi}+\mathcal{M}_{r\phi}}{\langle P \rangle} = \alpha_{\mathcal{R}} + \alpha_{\mathcal{M}},
    \label{eq:maxwell_alpha}
\end{equation}
where the subscripts refer to the Reynolds and Maxwell contributions, respectively. To diagnose the relative contributions of laminar and turbulent angular momentum transport at the dead--active zone interface, the Maxwell stress components are further decomposed into coherent and turbulent components via
\begin{align}
    4\pi\, \mathcal{M}_{r\phi}^{\text{coh}} &= -\langle{B_r}\rangle\langle{B_\phi}\rangle,	\label{eq:maxwell_stress_components}\\ 
    4\pi\,\mathcal{M}_{r\phi}^{\text{turb}} &= \langle{-B_rB_\phi}\rangle + \langle{B_r}\rangle\langle{B_\phi}\rangle, \nonumber \\
	4\pi\,\mathcal{M}_{\theta\phi}^{\text{coh}} &= -\langle{B_\theta}\rangle\langle{B_\phi}\rangle, \nonumber	\\
	4\pi\,\mathcal{M}_{\theta\phi}^{\text{turb}} &= \langle{-B_\theta B_\phi}\rangle + \langle{B_\theta}\rangle\langle{B_\phi}\rangle. \nonumber
\end{align} 
\indent Finally, the plasma-beta parameter, $\beta$, is defined as the ratio of the gas pressure to the magnetic pressure,
\begin{equation}
    \beta = \frac{8\pi P}{B^2},
    \label{equation:plasma_beta_parameter}
\end{equation}
where $B$ is the total magnetic field strength. The coherent vertical component is given by,
\begin{equation} 
    \beta_z^{\text{coh}} = \frac{8\pi\langle {P}\rangle}{\langle {B_z\rangle^2}},
    \label{eq:beta_vertical_coherent} 
\end{equation}
and used to quantify the strength of a large-scale vertical field.

\subsubsection{One-dimensional viscous-disc diffusion equation}
\label{section:one_dimensional_viscous_disc_diffusion_equation}
\indent An alternative and simplified approach to modelling accretion is through the one-dimensional viscous-disc diffusion equation, which neglects any wind-driven contribution and does not assume the source of viscosity is turbulence \citep{lynden-bell_evolution_1974}:
\begin{equation}
\frac{\partial \Sigma}{\partial t} = \frac{3}{R} \frac{\partial}{\partial R} \left[ \sqrt{R} \frac{\partial}{\partial R} \left( \bar{\nu} \Sigma \sqrt{R} \right) \right].
\label{equation:1d_viscous_diffusion_equation}
\end{equation}
Here the disc is assumed to be Keplerian, and $\bar{\nu}$ is the mean kinematic viscosity. Within the thin-disc approximation, $\bar{\nu}$ can be related to the $\alpha$-prescription via $\bar{\nu}\!\sim\!\alpha{c}_\text{s}H$. The corresponding average viscous torque, $\mathcal{G}$, and mass flux, $\mathcal{F}$, are given by,
\begin{align}
        \mathcal{G}(R,t) &= 3\pi\bar{\nu}R^{1/2}\Sigma, \label{eq:mathcalG}\\
        \mathcal{F}(R,t) &= -2R^{1/2}\partial_R{\mathcal{G}}\label{eq:mathcalF}.
\end{align}
To connect this to three-dimensional simulations, $\overline{\nu}$ is determined from a radial profile of $\alpha(R)$, averaged in azimuth and time, and vertically integrated over the disc.


\section{Results I. Magnetic Fields in the active zone and disc accretion}
\label{section:accretion_and_disc_structure}

In the first part of our results, we examine the evolution of the magnetic field in the MRI-active zone and its role in shaping the disc's accretion structure. We begin in Section~\ref{section:accretion_and_disc_structure_overview} with an overview of the magnetic field evolution across the entire disc, followed by a focused examination of the MRI-active zone in Section~\ref{section:magnetic_field_evolution_active_zone} and its vertical structure, in particular, in Section~\ref{section:vertical_accretion_structure}. Finally, Section~\ref{section:midplane_accretion_structure} investigates the midplane radial accretion structure. \\
\indent Unless stated otherwise, all results in this section are from \texttt{ZNF-FID} and averaged over the time interval $t_{\text{in}}\in[200,700]$, corresponding to approximately 100 local orbits at $R=0.75$ au, the primary midplane radial location of our analysis.
\begin{figure*}
    \centering
    \includegraphics[width=1.0\textwidth]{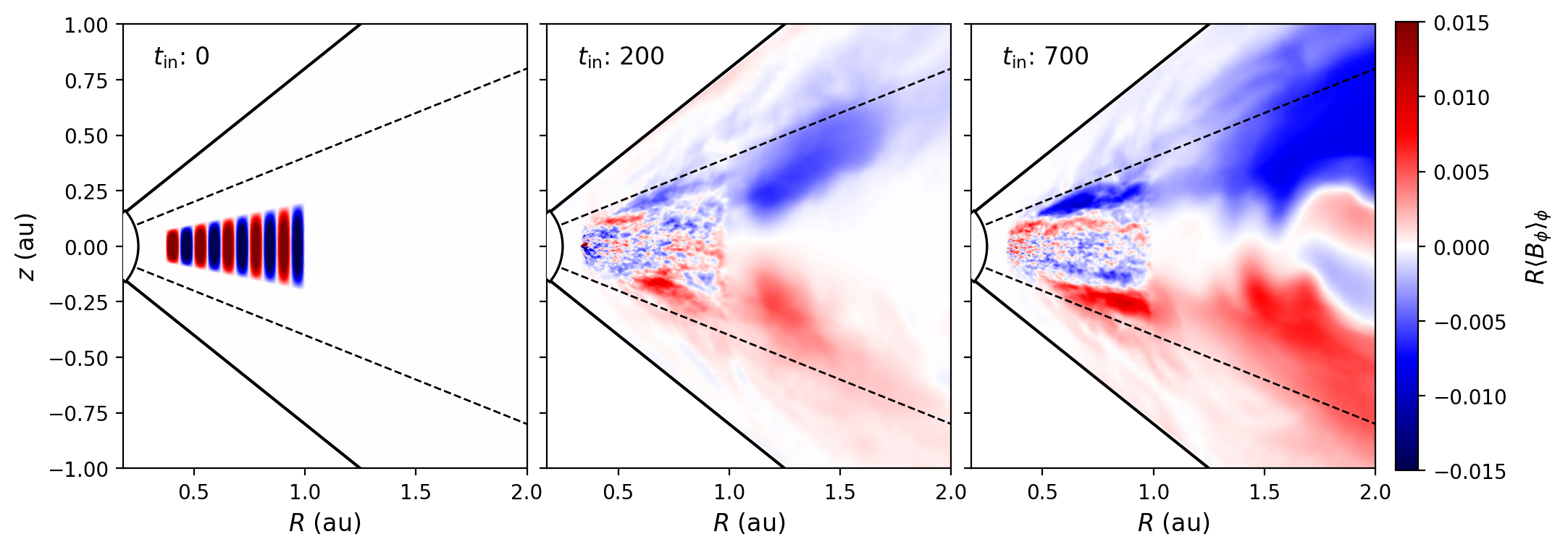}
    \caption{Meridional $(R,z)$ plots of the normalised, azimuthally averaged toroidal field, $R\langle{B_\phi}\rangle_{\phi}$, for three snapshots of \texttt{ZNF-FID}: $t_{\text{in}}=0$ (left), $t_{\text{in}}=200$ (middle) and $t_{\text{in}}=700$ (right). The dashed black lines denote the disc--corona transition at $z=\pm4H$, and the dead--active zone interface is at $R=1\,\text{au}$. Note the formation of characteristic MRI-driven turbulence in the inner disc, the presence of strong tightly wound magnetic field structures in the MRI-active disc surface layers (see Section \ref{section:vertical_accretion_structure}), and the complex toroidal magnetic field configuration that eventually permeates the entire dead zone (see Section \ref{section:magnetic_fields_dead_zone_overview}).}
    \label{fig:bx3_evolution}
\end{figure*}

\subsection{Overview of magnetic field evolution in the disc}
\label{section:accretion_and_disc_structure_overview}
The dynamics in the simulations are primarily governed by the MRI and the disc--corona temperature transition. We begin by presenting a rendering of the numerical domain at $ t_\text{in} = 30 $ for \texttt{ZNF-FID} in Fig.~\ref{fig:comparison_3D_plot}, which illustrates the rapid emergence of a self-consistently MRI-turbulent disc in the inner region ($ R < 1\,\mathrm{au} $), with MRI activity suppressed beyond the dead--active zone interface.\\
\indent To illustrate the overall evolution of magnetic structures in the disc, we present snapshots of the normalised, azimuthally averaged toroidal field, $R\langle B_\phi \rangle_{\phi}$, in the meridional $(R,z)$ plane, from the initial condition through to $t_\text{in}=700$, in Fig.~\ref{fig:bx3_evolution}. Radially, a turbulent MRI-active region emerges, with the MRI suppressed in the dead zone. This suppression results from the choice of magnetic Reynolds number (right panel of Fig.~\ref{fig:non_ideal_MHD_setup}), which lies below the empirical threshold for stratified ZNF MRI stability, $\text{R}_\text{m} \lesssim 3000$ \citep{flock_turbulence_2012}, effectively acting as a numerical switch for MRI activity. Over time, a complex toroidal field morphology emerges and gradually extends into the dead zone, a process examined in more detail in Section~\ref{section:magnetic_field_structures}. Meanwhile, in the vertical direction, strong toroidal magnetic field structures build up at the disc--corona temperature transition in the MRI-active zone.  This configuration exerts significant magnetic torques on the disc, driving localised radial surface-layer accretion (see Section~\ref{section:vertical_accretion_structure}). Finally, large-scale, coherent poloidal magnetic field loops are generated in the ideal-MHD region. These span the radial and meridional extent of the MRI-active zone and are explored further in Section \ref{section:large_scale_coherent_poloidal_magnetic_field_loops}.

\subsection{Magnetic field evolution in the active zone}
\label{section:magnetic_field_evolution_active_zone}

\subsubsection{Emergence and saturation of the MRI}

\begin{figure}
    \includegraphics[width=1\columnwidth]{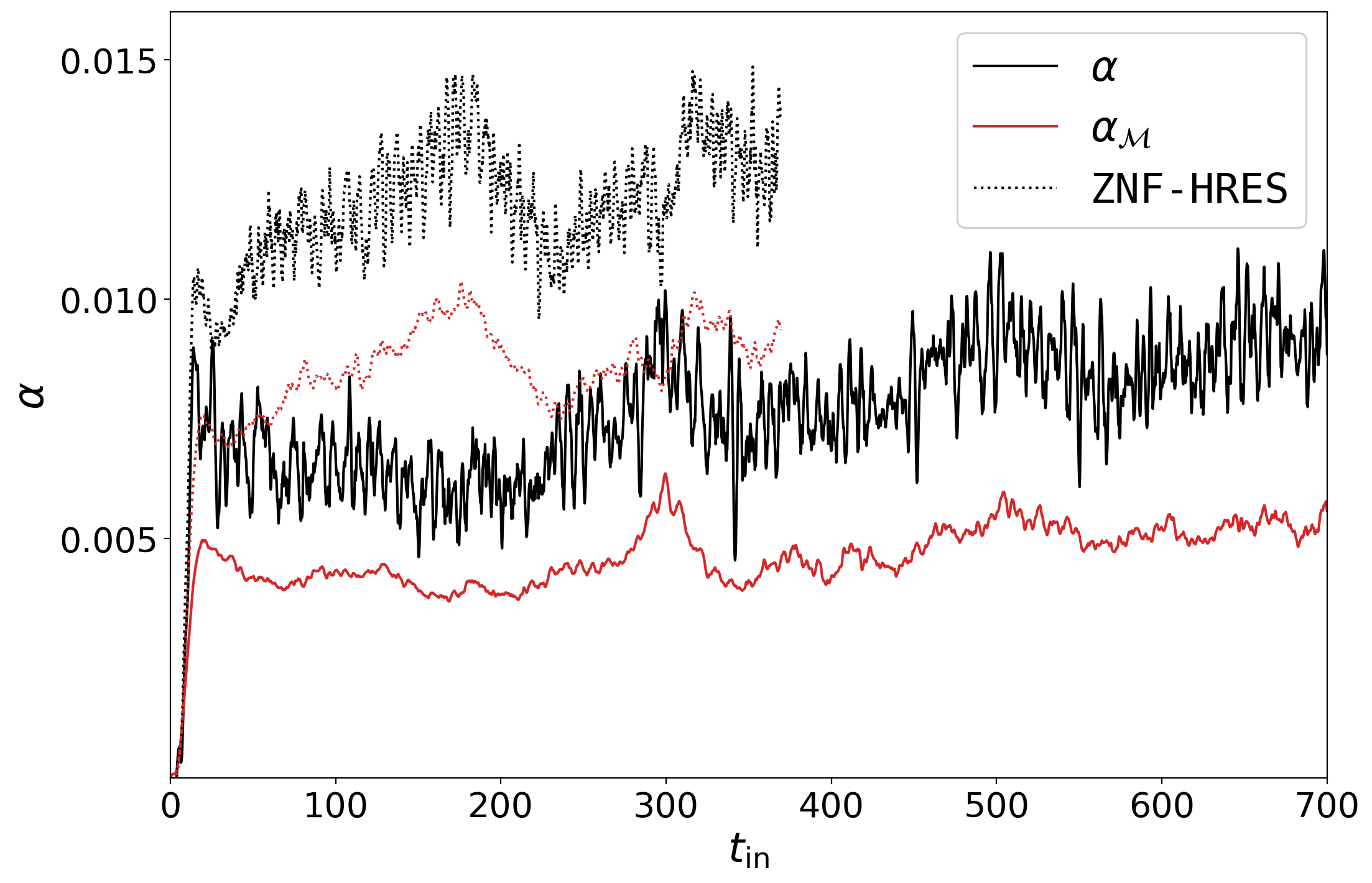}
    \caption{Temporal evolution of the volume-averaged total $\alpha$ (black) and Maxwell $\alpha_\mathcal{M}$ (red) in the MRI-active zone, which are larger in the higher-resolution simulation \texttt{ZNF-HRES} (dotted) compared to \texttt{ZNF-FID} (solid). The volume average is computed over the spherical wedge: $r\in[0.5,1]\,\text{au}$, $\theta \in[z=-4H, z=4H]$ and $\phi\in[0,2\pi)$.}    \label{fig:alpha_volume_average_active_resolution_comparison}
\end{figure}

We begin by characterising the saturated MRI dynamo in the ideal-MHD region of the disc. As shown in Fig.~\ref{fig:alpha_volume_average_active_resolution_comparison}, the volume-averaged $\alpha$ in the ideal-MHD zone exhibits rapid exponential growth, reaching saturation within approximately $3\,t_{\mathrm{DZI}}$, and remains sustained for the duration of the simulation. The time-averaged value is $\alpha\!\sim\!8\times10^{-3}$, with the Maxwell component contributing $\alpha_{\mathcal{M}}\!\sim\!5\times10^{-3}$. This behaviour confirms the presence of a persistent MRI-driven dynamo operating via a fully developed turbulent state. For comparison, the global ZNF dead--active zone interface simulations with radiative transfer by \citet{flock_3d_2017} report a higher volume-averaged stress of $\alpha\!\sim\!3\times10^{-2}$ in the MRI-active zone of their fiducial model. However, their azimuthally restricted domain employed a higher effective resolution, equivalent to 2048 cells in azimuth. \\
\indent 
To assess the sensitivity of this evolution to the numerical grid, we conduct a resolution test using \texttt{ZNF-AZ}. The disc structure, evolution, and large-scale magnetic features resemble those in \texttt{ZNF-FID}, but differences arise due to changes in the saturated strength of the ZNF MRI. As shown in Fig. \ref{fig:alpha_volume_average_active_resolution_comparison}, the total and Maxwell components of the saturated, volume-averaged $\alpha$ in the active zone, are about $60$ per cent higher in \texttt{ZNF-HRES} (dotted lines), enhancing angular momentum transport and accelerating viscous evolution by a similar factor, under thin-disc assumptions. This underscores the difficulty of extracting precise accretion rates from global ZNF MRI simulations at these resolutions. Whether higher-resolution global simulations reproduce the non-converging behaviour in stratified isothermal shearing boxes \citep[see table 1 in][]{ryan_resolution_2017}, where $\alpha$ falls with increasing resolution, remains unclear.

\subsubsection{Large-scale coherent poloidal magnetic field loops}
\label{section:large_scale_coherent_poloidal_magnetic_field_loops}

\begin{figure}
    \includegraphics[width=0.98\columnwidth]{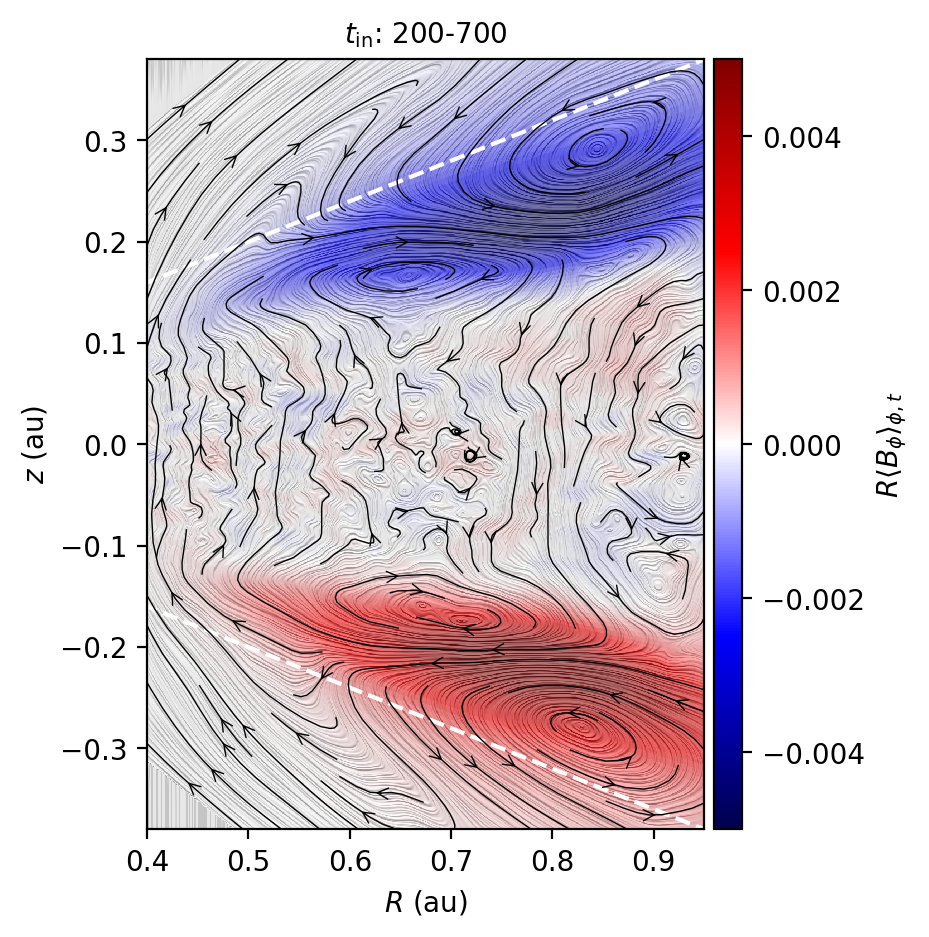}
    \caption{Magnetic field configuration for \texttt{ZNF-FID} in the active zone, averaged in azimuth and over the interval $t_{\text{in}}\in[200,700]$ to enhance the visibility of coherent field structures. The poloidal magnetic field, $\langle\mathbf{B_p}\rangle_{\phi,t}$, shown using black field lines and a LIC overlay in greyscale, forms coherent clockwise loops spanning the extent of the MRI-active region. The sign of the normalised toroidal field, $R\langle B_\phi\rangle_{\phi,t}$, (background colour) near the disc surface layers (dashed white lines), is negatively correlated with the sign of the radial magnetic field.}
    \label{fig:BpLIC_BPhi_small_600700}
\end{figure}

In the MRI-active region, large-scale poloidal field loops are generated self-consistently from an initial toroidal field, which is defined in equation~\eqref{eq:magnetic_initial_condition}, and grow to fill the extent of the active disc region. Fig.~\ref{fig:BpLIC_BPhi_small_600700} shows the magnetic field configuration in the inner disc, averaged in azimuth and over the time interval $t_{\mathrm{in}}\in[200,700]$, to enhance the visibility of the coherent field structures against the background turbulent state.  The poloidal field configuration, $\langle\mathbf{B_p}\rangle_{\phi,t}$, is visualised using field lines (black) and a LIC overlay (greyscale). A set of concentric, clockwise, large-scale poloidal field loops is clearly visible. Once formed, these structures persist throughout the extent of the simulation (see Section~\ref{section:magnetic_field_structures}), and their comparable evolution in the azimuthally restricted, long-runtime simulation, \texttt{ZNF-LONG}, indicates that a full azimuthal domain is not required to capture this process.\\
\indent Previous work by \citet{jacquemin-ide_magnetorotational_2024} demonstrated the self-consistent generation of large-scale poloidal magnetic field loops in global MRI-active simulations of black hole accretion discs. In contrast to our initial conditions (see Section~\ref{section:initial_conditions}), their study began with a strong, uniform axisymmetric toroidal magnetic field with a plasma-beta of $ \beta \!=\! 5 $ and a scale height of $ H/r \!\sim\! 0.2 $ in the bulk disc, corresponding to a relatively thick accretion torus structure. However, they raised concerns regarding the efficiency of dynamo-driven field generation in thinner discs, which are less effective at advecting magnetic flux \citep{lubow_magnetic_1994}. Therefore, this work provides evidence for the formation and persistence of similar large-scale poloidal field loops in a thinner disc with $ H/R = 0.1 $, a regime more representative of protoplanetary discs. This lends tentative support to the notion that thin MRI-active discs can generate and sustain large-scale magnetic fields self-consistently, without requiring an initial vertical-net flux.

\subsection{Vertical structure of the active zone}

\label{section:vertical_accretion_structure}
\subsubsection{Surface-layer accretion}
\label{section:surface_layer_accretion}
In the MRI-active, ideal-MHD region of the disc a vertically layered accretion structure develops, and is sustained, in all the simulations, including the pure active model \texttt{ZNF-AZ}. To our knowledge, this feature has not previously been reported in the ZNF configuration. The majority of the accretion is localised just below the disc surface layers at $z\approx \pm3H$, in a near symmetric manner across the disc midplane. To quantify the connection between accretion and magnetic stresses, we perform an extensive temporal average over the interval $t_{\mathrm{in}}\in[200,700]$ in this section. \\ 
\begin{figure}
    \centering  
    \includegraphics[width=0.98\columnwidth]{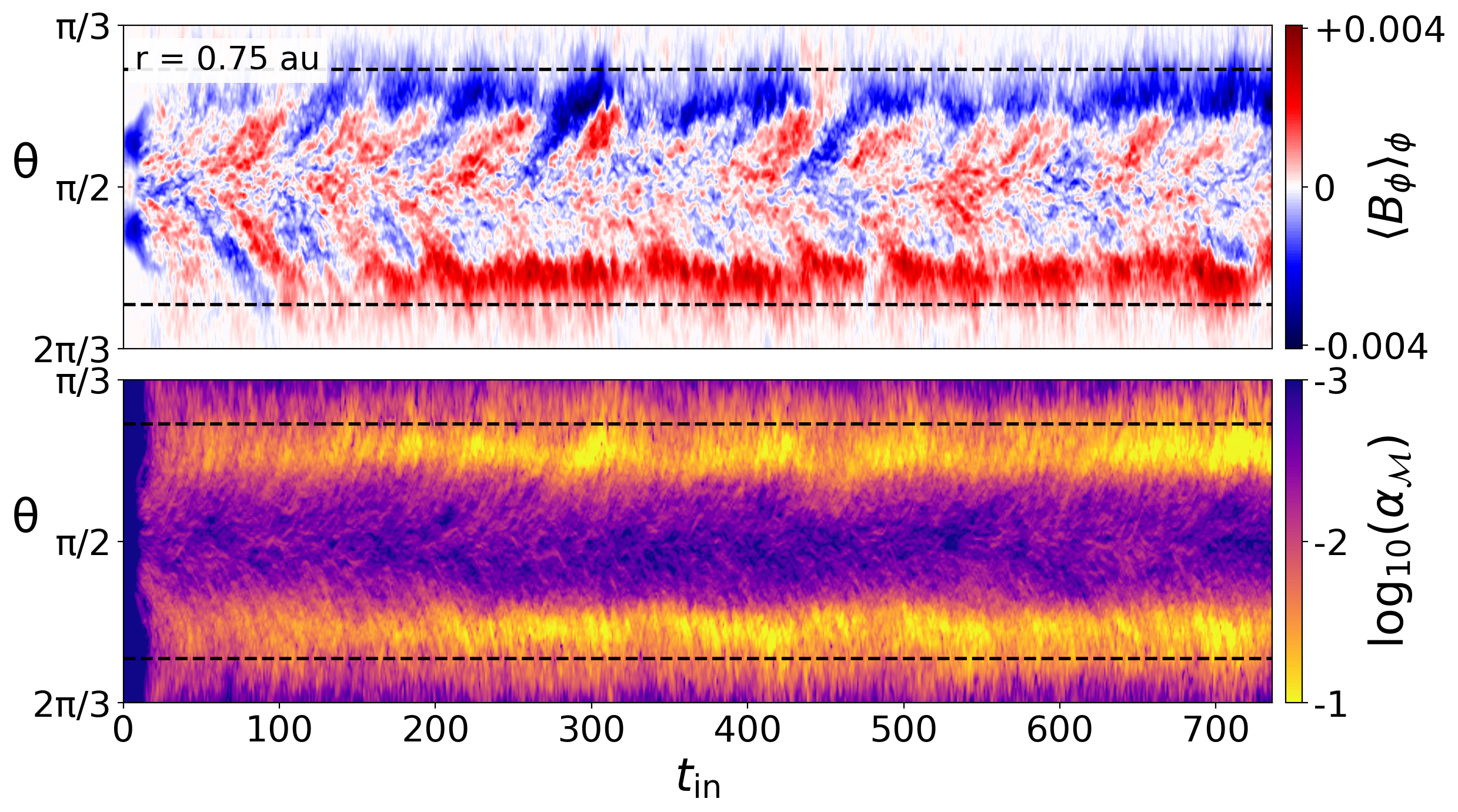} \\
    \caption{Spacetime $(\theta,t)$ diagrams for \texttt{ZNF-FID}, along the contour $r=0.75$ au, showing the azimuthally averaged toroidal field, $\langle B_\phi \rangle_\phi$ (top), and Maxwell $\alpha$ component \eqref{eq:maxwell_alpha} (bottom). The disc--corona transition is at $z=\pm4H$ (dashed black lines) and the surface layers above the MRI-active disc have a tightly wound magnetic field configuration that drives strong local radial angular momentum transport.}
    \label{fig:theta_spacetime_r0.75_fid.png}
\end{figure}
\begin{figure}
    \centering  
    \includegraphics[width=0.98\columnwidth]{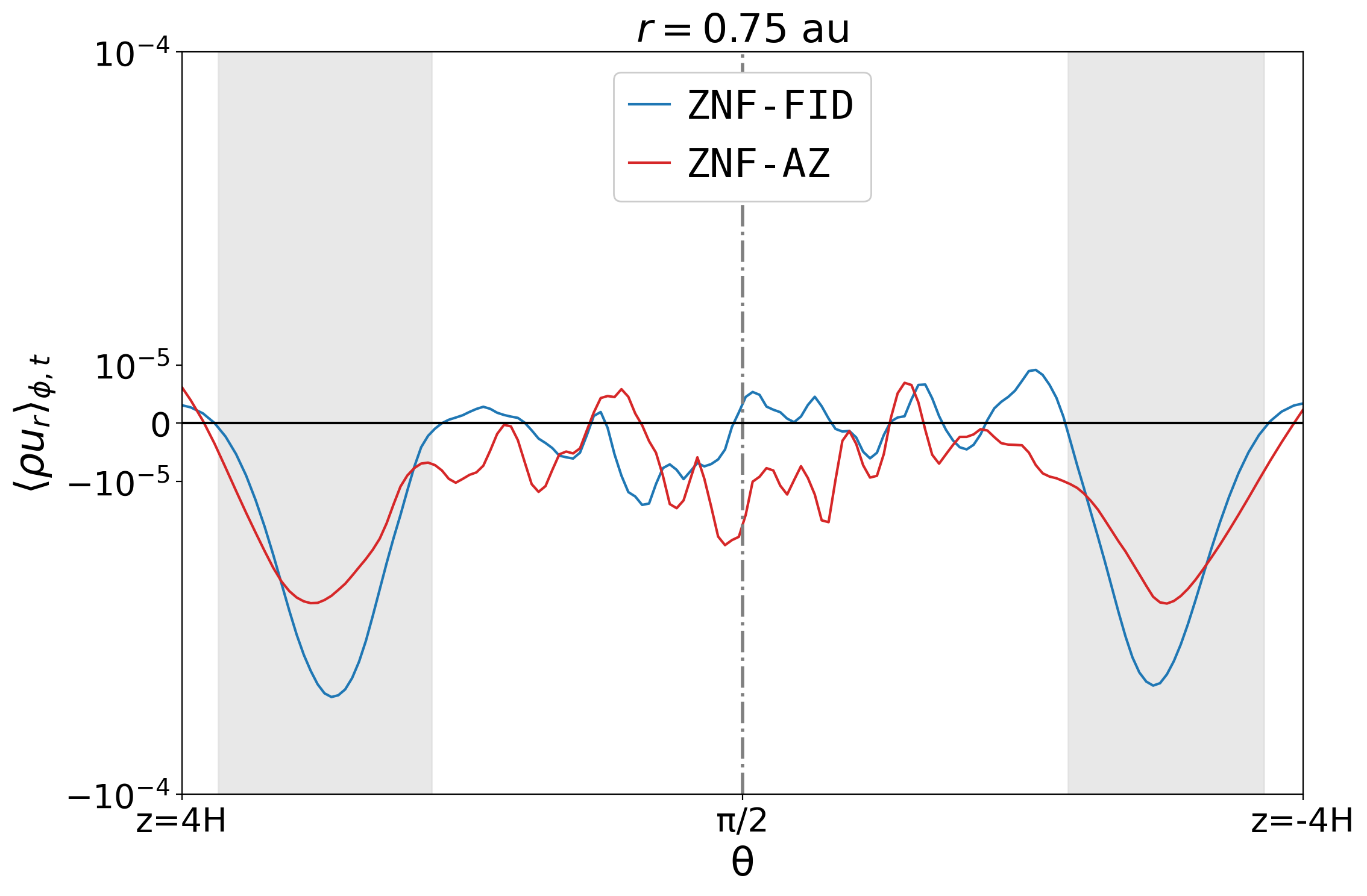}
    \caption{Vertical structure of the accretion flow in the MRI-active region for \texttt{ZNF-FID} (blue) and \texttt{ZNF-AZ} (red), quantified by $\langle\rho u_r\rangle_{\phi,t}$. The plot is taken over the spherical contour $r=0.75\,\text{au}$ and has been averaged in azimuth and over the interval $t_{\text{in}}\in[200,700]$. There are strong localised accretion flows (grey-shaded regions) in the disc surface layers, which are associated with radial angular momentum transport of a magnetic origin, quantified in Fig.~\ref{fig:stress_components_r0.75_fid.png}.}
    \label{fig:rhour_r0.75_fid.png}
\end{figure}
\indent To understand the vertical morphology in the active zone, we examine the magnetic field structures and their associated stresses over the spherical contour $r=0.75$ au. The $(\theta,t)$ spacetime diagram presented in Fig. \ref{fig:theta_spacetime_r0.75_fid.png} exhibits the operation of the stratified MRI dynamo close to the midplane throughout the entire simulation, showing the well-known butterfly pattern in $\langle B_\phi\rangle_\phi$, which corresponds to periodic azimuthal-field cycles \citep[e.g.][]{brandenburg_dynamo-generated_1995}. \\
\indent However, close to the disc surface layers there is a different phenomenon, responsible for driving surface-layer localised accretion. The strong build-up of radial and toroidal field dramatically alters the vertical structure of magnetic fields in the active region. First, the corresponding plasma-$\beta$ parameter, defined in \eqref{equation:plasma_beta_parameter}, along this contour, varies from $\sim\!10^4$ at the midplane, down to $\sim\!10$ at $z=\pm3H$. Whilst this is not sufficient to suppress the MRI through magnetic tension, it indicates the presence of two different magnetic field strength regimes. Meanwhile, local magnetic field correlations drive a sustained $\alpha_\mathcal{M}\!\sim\!10^{-1}$ in the disc surface layers, in sharp contrast to the weak $\alpha_\mathcal{M}\!\sim\!10^{-3}$ at the midplane, as shown in Fig. \ref{fig:theta_spacetime_r0.75_fid.png}. \\
\indent In Fig. \ref{fig:rhour_r0.75_fid.png}, we present the resulting vertically layered accretion profile, characterised by
$\langle \rho u_r \rangle_{\phi,t}$, along the same spherical contour for both \texttt{ZNF-FID} and \texttt{ZNF-AZ}. The averaged accretion rate is an order of magnitude stronger (grey-shaded regions) in the surface layers compared to the midplane in \texttt{ZNF-FID}. Whilst this accretion flow is not supersonic, it reaches approximately 75 per cent of the local disc sound speed, $c_\text{s} = \varepsilon R^{-1/2}$, at the altitude of the maximal accretion rate. This bears resemblance to the elevated accretion structures observed in vertical-net flux global MRI-active simulations performed by \citet{zhu_global_2018} and \citet{jacquemin-ide_magnetic_2021}. \\
\indent Finally, we decompose the vertical profiles of these stresses further in Fig. \ref{fig:stress_components_r0.75_fid.png} to take account of the turbulent and coherent components, defined in equation \eqref{eq:maxwell_stress_components}. Even though vertical stresses contribute $\sim\!R/H$ more to accretion than vertically integrated radial stresses, under the thin-disc approximation, the dominant driver of outward angular momentum transport is the turbulent $r\phi$ component of the Maxwell stress, $\mathcal{M}^{\text{turb}}_{r\phi}$. Therefore, despite the large-scale, laminar looking magnetic field structures (see Fig. \ref{fig:bx3_evolution}), the vertical accretion structure in the MRI-active zone is being driven by significant vertical inhomogeneity in the radial torque imposed by $\mathcal{M}^{\text{turb}}_{r\phi}$. \\
\begin{figure}
    \centering  
    \includegraphics[width=0.98\columnwidth]{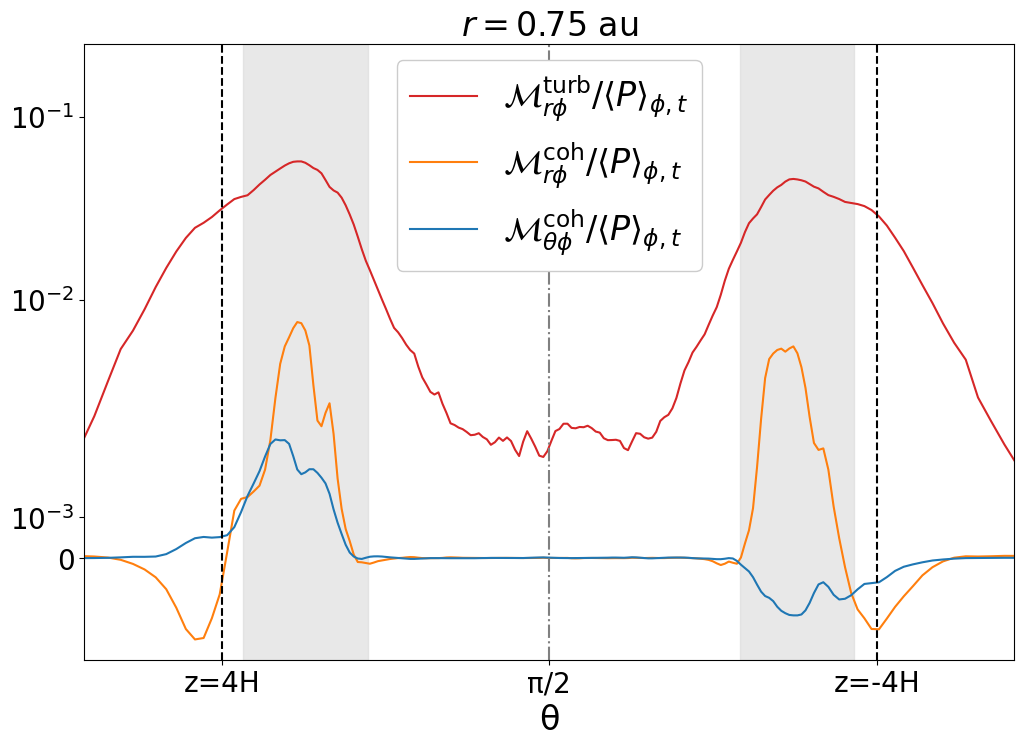} \\
    \caption{Decomposition of the total Maxwell stress components, defined in equation \eqref{eq:maxwell_stress_components}, along the spherical contour $r=0.75\,\text{au}$. The stresses are normalised by the local pressure to enable quantitative comparisons across altitudes, and have been averaged in azimuth and over the interval $t_{\text{in}}\in[200,700]$. The grey-shaded regions indicate the localised surface accretion flows at $z\approx \pm 3H$, shown in Fig.~\ref{fig:rhour_r0.75_fid.png}.}
    \label{fig:stress_components_r0.75_fid.png}
\end{figure}
\indent To test the sensitivity of surface-layer accretion in the active region to the numerical boundary proximity and the dead--active zone interface, we conducted an additional simulation, \texttt{ZNF-AZ} (see Table \ref{table:list_global_simulations}), where the non-ideal MHD diffusivity profiles are removed, rendering the entire disc fully MRI turbulent. In this model, there is also the accumulation of a strong, tightly wound toroidal magnetic field just beneath the disc's surface layers, as shown in the $(\theta, t)$ spacetime diagrams in Fig. \ref{fig:theta_spacetime_r0.75_az.png}. This should be compared with the corresponding contour from the \texttt{ZNF-FID} model in Fig. \ref{fig:theta_spacetime_r0.75_fid.png}. Moreover, the configuration extends across the full radial extent of the disc, including regions well beyond the inner boundary. This suggests that the behaviour is not solely driven by boundary conditions or the imposed dead--active zone interface, but is intrinsic to the ZNF MRI and thermodynamic setup used. Further work, in a more controlled environment, is needed to determine the exact mechanism that forms this magnetic field morphology. For now, we hypothesise the following general argument. 
\begin{figure}
    \centering  
    \includegraphics[width=0.98\columnwidth]{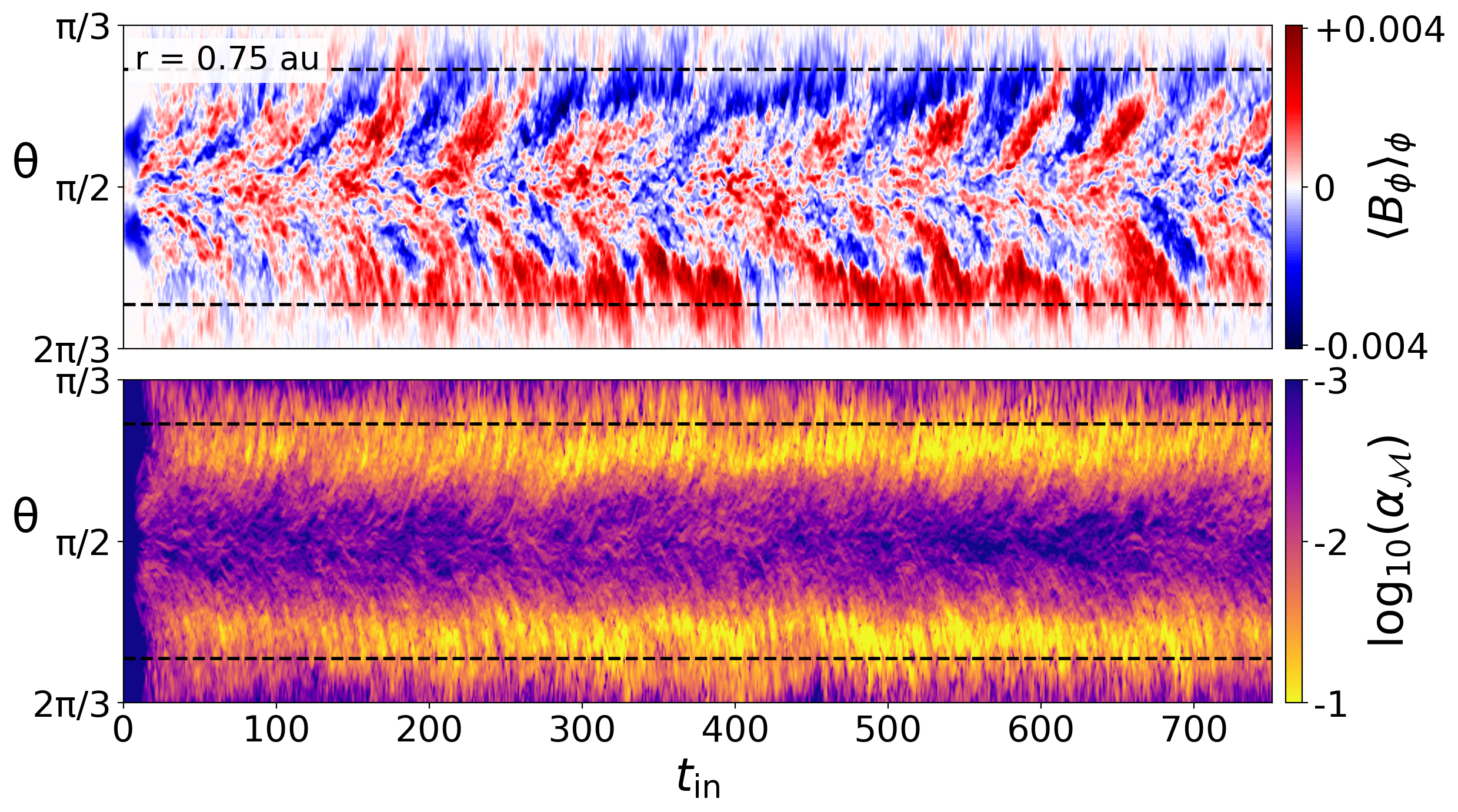} \\
    \caption{Spacetime $(\theta,t)$ diagrams for the fully MRI-active disc simulation, \texttt{ZNF-AZ}, along the contour $r=0.75$ au for the azimuthally averaged toroidal field, $\langle B_\phi \rangle_\phi$ (top), and Maxwell $\alpha$ component (bottom). The disc--corona temperature transition is at $z=\pm4H$ (dashed black lines) and these results should be compared with equivalent plots for \texttt{ZNF-FID} presented in Fig. \ref{fig:theta_spacetime_r0.75_fid.png}.}
    \label{fig:theta_spacetime_r0.75_az.png}
\end{figure}

\subsubsection{Surface-layer accretion mechanism}
\indent The sharp transition to a hot corona, defined in equation \eqref{equation:target_temperature}, alters the buoyancy forces that push the dynamo-generated magnetic flux away from the disc midplane. In the simplified one-dimensional flux-tube model of \cite{torkelsson_magnetic_1993}, a sharp gradient change in the vertical pressure profile can stall buoyantly rising azimuthal flux tubes of fixed length at the disc--corona transition, resulting in the build-up of these magnetic flux tubes just below the disc surface layers. \\
\indent In our simulations, the large-scale poloidal magnetic field loops fill the active zone, but do not expand vertically into the corona. Instead, the generated large-scale radial field is halted at the disc--corona transition, as evidenced by the bunching of radial flux at $|z|\approx0.2\,\text{au}$ in Fig.~\ref{fig:BpLIC_BPhi_small_600700}. We attribute this behaviour to the modified buoyancy profile at the transition, but note that it represents a more complex configuration than the simplified model of \citet{torkelsson_magnetic_1993} discussed above. The $\Omega$-effect acts on the buoyantly trapped radial field, accessing the disc's shear reservoir to generate strong toroidal fields at both surface layers, as supported by the observed anti-correlation between the signs of $B_r$ and $B_\phi$ in Fig.~\ref{fig:BpLIC_BPhi_small_600700}. Once formed, the large-scale poloidal loops (which are always clockwise in our simulations) stabilise a particular polarity, effectively locking in the sign of $B_\phi$ produced by the $\Omega$-effect for the remainder of the evolution. \\
\indent Additional evidence for the influence of a hot corona on the structure of MRI-active discs is found in previous studies of both ZNF and vertical-net flux simulations. For example, the shearing-box simulations by \cite{bambic_local_2024}, with a self-consistently computed two-temperature disc--corona structure demonstrate similar features (see bottom panels of their fig. 1). Comparable behaviour is also seen in global simulations of vertical-net flux presented in the second paper of this series and in ZNF simulations by J. Jacquemin-Ide (private communication). In contrast, the global ZNF MRI-active simulations by \citet{parkin_global_2013} show no such evidence, despite implementing a disc--corona temperature transition (see their fig. 7). However, their method imposes the transition using a different, possibly less sharp, technique and employs a geometrically thinner disc ($\varepsilon = 0.05$), a regime that has previously raised concerns about the generation of large-scale poloidal fields from the MRI dynamo \citep{jacquemin-ide_magnetorotational_2024}.
\subsection{Midplane accretion structure}
\label{section:midplane_accretion_structure}
\begin{figure}
    \includegraphics[width=0.98\columnwidth]{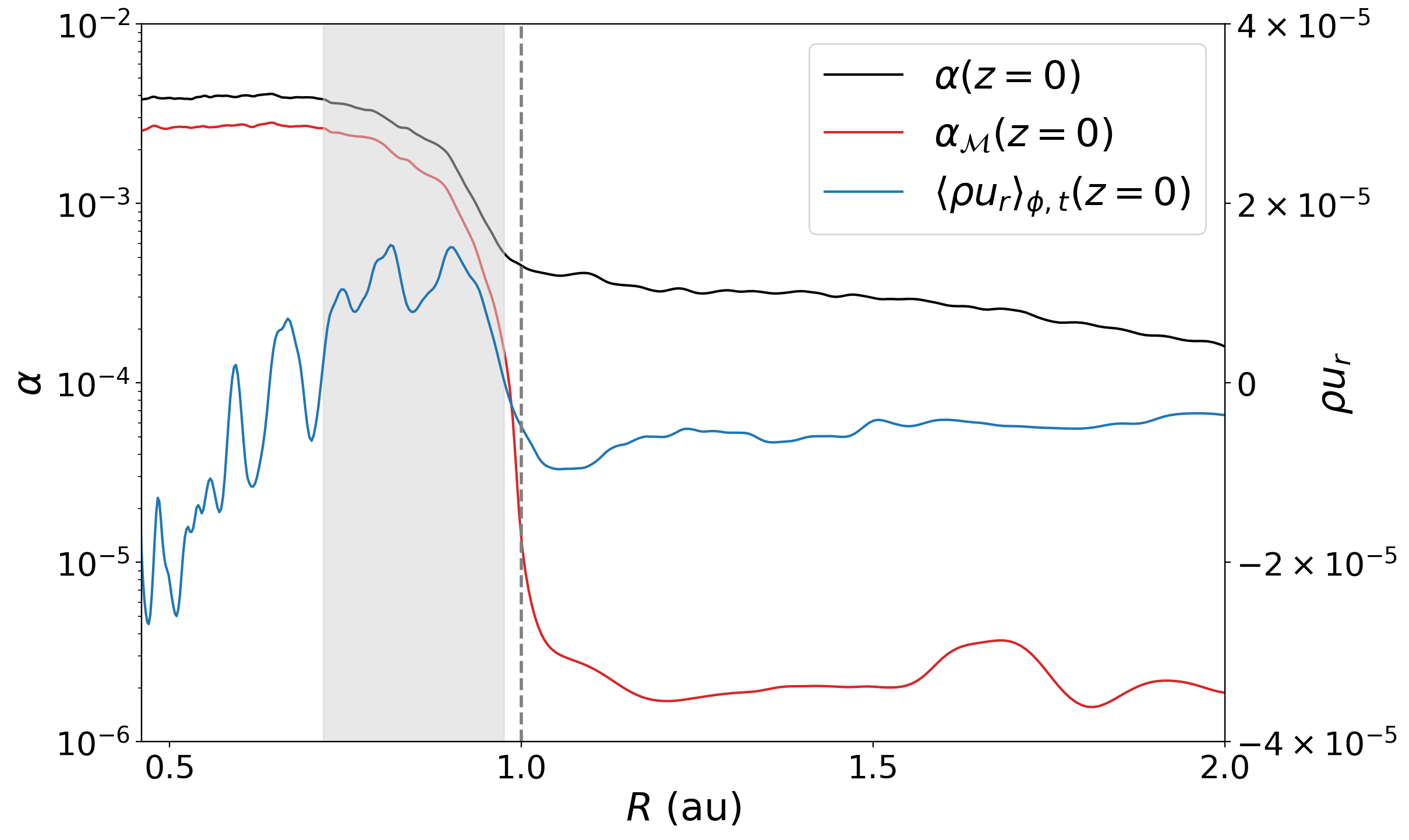}
    \caption{Radial profiles of $\alpha$ (black), $\alpha_\mathcal{M}$ (red), and the accretion flow, quantified by $\langle\rho u_r\rangle_{\phi,t}$ (blue). The quantities are evaluated at the midplane and have been averaged in azimuth and over the interval $t_{\text{in}}\in[200,700]$. The grey-shaded region highlights that the midplane flow is outward at the outer edge of the MRI-active zone.}
    \label{fig:test10_midplane_averages_800_2800}
\end{figure}

The accretion flow at the midplane is controlled by the stress profiles in the disc. Fig. \ref{fig:test10_midplane_averages_800_2800} shows radial profiles of $\alpha$ and $\alpha_{\mathcal{M}}$ at the midplane. In the inner MRI-active zone,  $\alpha\!\sim\!3\times{10}^{-3}$ with $\alpha_{\mathcal{M}}/\alpha_{\mathcal{R}}\!\sim\!3$, consistent with typical MRI-saturated dynamo behaviour. Furthermore, $\alpha_\mathcal{M}$ at the midplane is roughly $40$ per cent lower compared to the volume-averaged value over the MRI-active region shown in Fig. \ref{fig:alpha_volume_average_active_resolution_comparison}. This is the result of the inhomogeneous vertical magnetic stress structure in the active zone, shown in Fig.~\ref{fig:stress_components_r0.75_fid.png}. \\
\indent Meanwhile, in the inner dead zone, the total midplane $\alpha$ is $\sim\!5\times10^{-4}$, which is nearly an order of magnitude lower than in the active zone. This stress is driven almost entirely by the Reynolds component, attributed to large-scale density waves emanating from vortex structures (see Section~\ref{section:vortices}; \citealt{chametla_turbulent_2024}), fluctuations at the dead--active zone interface \citep{faure_thermodynamics_2014}, and the MRI-active zone \citep{heinemann_excitation_2009}.\footnote{Section~\ref{section:magnetic_fields_poloidal_field} extends this analysis to the vertical Maxwell stress profile in the inner dead zone, driven by large-scale magnetic field structures.} These waves decay with radius due to dissipation by weak shocks \citep[e.g.][]{faure_thermodynamics_2014}, leading to a decrease in the associated Reynolds $\alpha$. \\
\indent The midplane accretion flow is quantified with $\langle\rho u_r\rangle_{\phi,t}$, and its radial profile is shown in Fig. \ref{fig:test10_midplane_averages_800_2800}. In the inner MRI-active region ($R\lesssim0.75$ au) and throughout the dead zone, the time-averaged flow is inward. However, at the outer edge of the MRI-active region $(0.75\,\text{au}\lesssim R \lesssim 0.95\,\text{au})$ there is sustained outward flow at the midplane. This configuration is present in all four dead--active zone interface simulations and persists throughout their evolution. We attribute this outward flux to the radial gradient of the torque distribution, noting that the direction of radial mass flux is set by the sign of $-\partial_R(R^2\alpha P)$, in the absence of strong vertical stresses \citep[see][]{lesur_magnetohydrodynamics_2021}. \\
\indent Finally, we stress the physical significance of this flow structure. The passage of highly turbulent, hot material towards and through the dead--active zone interface modifies dust aggregation within putative traps -- pressure maxima and vortices -- and may also affect the advective transport of large-scale poloidal magnetic flux in the vertical-net flux configuration.

\section{Results II. large-scale Hydrodynamic Structures}
\label{section:large_scale_hydrodynamic_structures}

We continue our analysis by examining the formation, spatial morphology, and evolution of the large-scale HD structures that form at the dead--active zone interface. We investigate the axisymmetric pressure structure in Section~\ref{section:pressure_bump}, followed by the anticyclonic vortices in Section~\ref{section:vortices}. We highlight the rapid and robust nature of their formation, especially within the broader temporal variability of the inner disc.
\subsection{Pressure bump}
\label{section:pressure_bump}
\begin{figure}
    \subfloat{
        \includegraphics[width=0.98\columnwidth]{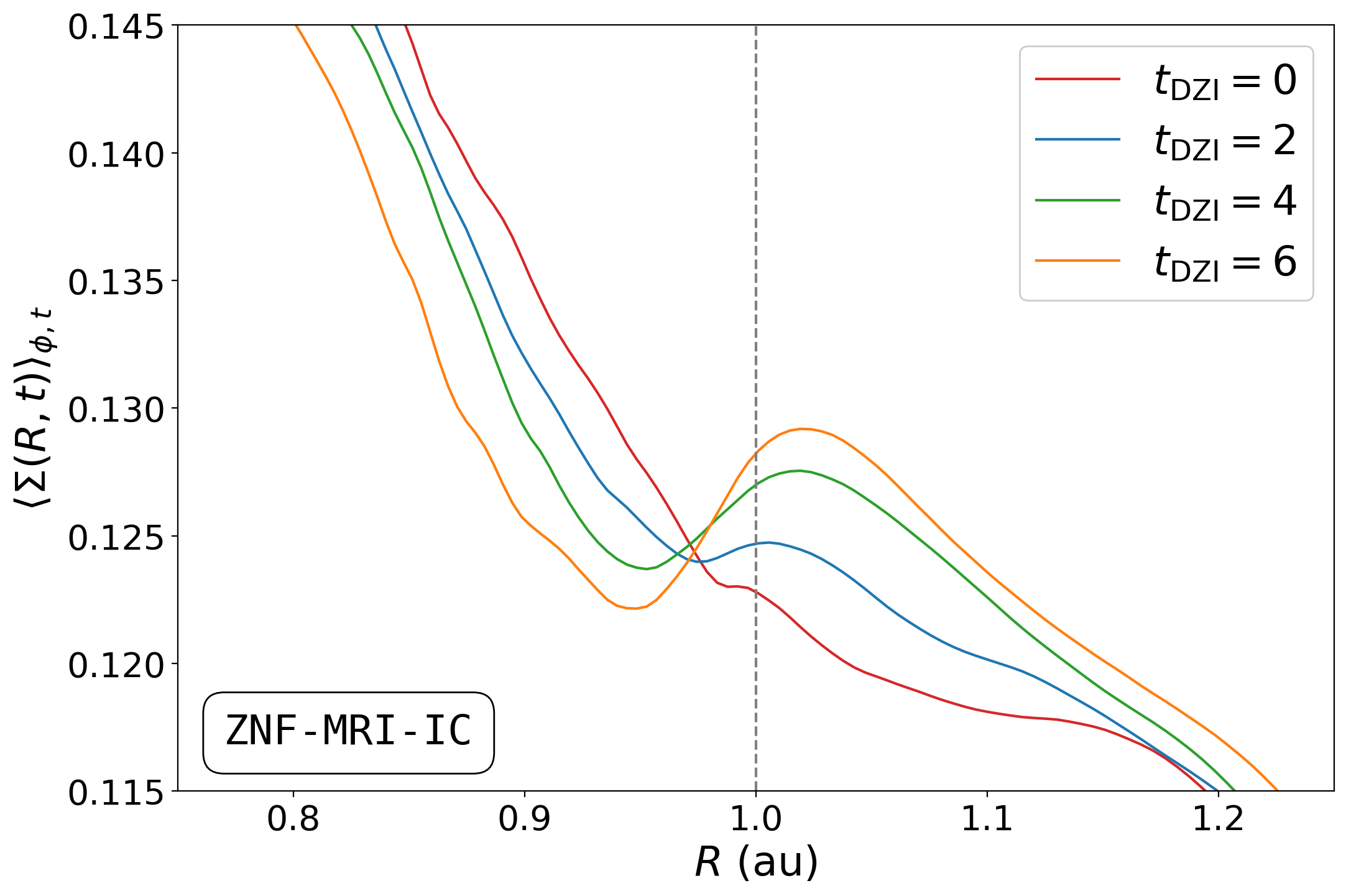}
    }
    \\
    \subfloat{
        \includegraphics[width=0.98\columnwidth]
        {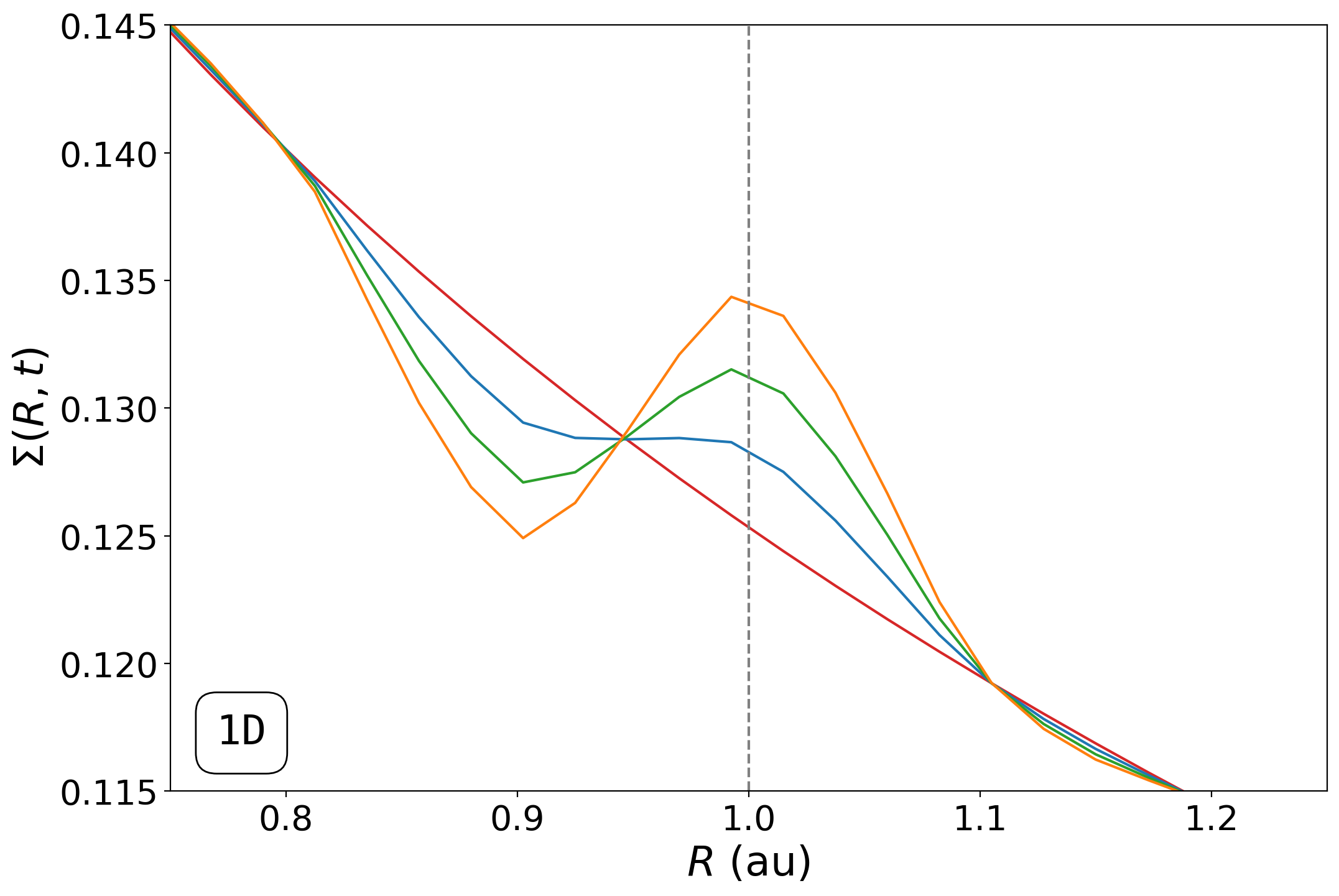}
    }
    \caption{Radial profiles of the surface density, $\Sigma$, over the initial six orbits ($t_\text{in}=48$) at the dead--active zone interface (dashed grey line). Note the similarity in evolution between the three-dimensional MHD simulation \texttt{ZNF-MRI-IC} (top) and the one-dimensional viscous HD model (bottom), and the density enhancement around the dead--active zone interface. In the top panel, $t_\text{DZI}=0$ references the time that the non-ideal MHD model is placed on top of a fully MRI-turbulent disc, and $\Sigma$ is smoothed over a small temporal average of $t_{\text{in}}=0.5$.}
    \label{fig:sigma_evolution_compare_1D_versus_3D}
\end{figure}

\begin{figure}
    \includegraphics[width=0.98\columnwidth]{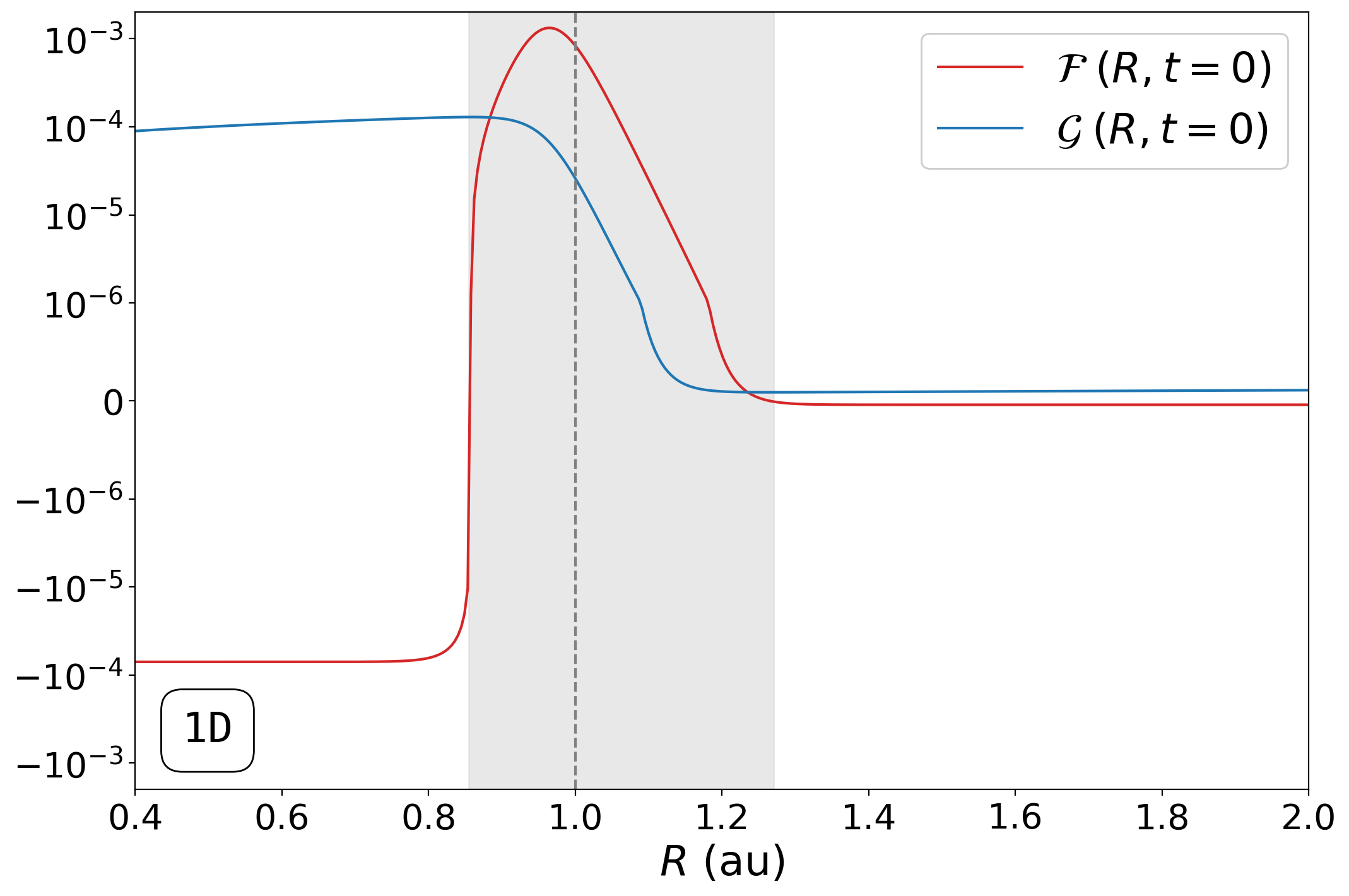}
    \caption{Radial profiles of the initial average viscous torque, $\mathcal{G}$ (blue), and mass flux, $\mathcal{F}$ (red), for the one-dimensional viscous-disc diffusion model, outlined in Section~\ref{section:pressure_bump}. The grey-shaded region denotes where the mass flux is outward $(\mathcal{F}>0)$, explaining the formation of a local mass enhancement just outside the dead--active zone interface (dashed grey line) shown in Fig. \ref{fig:sigma_evolution_compare_1D_versus_3D}.}
    \label{fig:1D_diffusion_model_F_G_comparison}
\end{figure}

A coherent, local axisymmetric density maximum forms just outside the dead--active zone interface in \texttt{ZNF-FID}, developing rapidly within a few local orbits. To test any dependence of this rapid evolution on the initial conditions, we perform the \texttt{ZNF-MRI-IC} simulation, in which the non-ideal MHD profiles are imposed on top of a fully MRI-turbulent disc. The pressure bump still forms in a comparable manner after these non-ideal terms are introduced, indicating that this outcome is not sensitive to the specific initial setup. \\
\indent Radial profiles of the surface density, $\Sigma(R)$, during the early stages of evolution are shown in the top panel of Fig.~\ref{fig:sigma_evolution_compare_1D_versus_3D}. These profiles demonstrate that the local maximum forms due to a density enhancement at the inner edge of the dead zone, rather than solely from the inward draining of material in the MRI-active zone. \\
\indent To investigate this process, we used the one-dimensional viscous-disc diffusion formalism introduced in Section \ref{section:one_dimensional_viscous_disc_diffusion_equation}. The setup assumed an initial Keplerian configuration, as given in equation \eqref{equation:1d_viscous_diffusion_equation}, and an initial surface density of $\Sigma \propto R^{-1/2}$, consistent with the initial condition \eqref{eq:density_initial_condition}. Fixed radial boundary conditions were applied to mimic the buffer zones used in the three-dimensional simulations. The formalism was closed by setting a vertically integrated radial profile of $\alpha_{\mathcal{M}}$, consistent with the midplane profile shown in Fig.~\ref{fig:test10_midplane_averages_800_2800}, thereby defining a steady profile for the kinematic viscosity, $\bar{\nu}$. We then solved the system on a radial grid comparable to that used in \texttt{ZNF-MRI-IC}; the resulting evolution of $\Sigma$ is shown in the bottom panel of Fig.~\ref{fig:sigma_evolution_compare_1D_versus_3D}. \\
\indent The one-dimensional model closely reproduces the evolution observed in \texttt{ZNF-MRI-IC} over the first six local orbits, confirming that the density enhancement at the dead--active zone interface results from a radial viscous process. To further quantify this behaviour, Fig.~\ref{fig:1D_diffusion_model_F_G_comparison} shows the initial radial profiles of the average viscous torque, $ \mathcal{G} $, and the corresponding mass flux, $ \mathcal{F} $, as defined in equations \eqref{eq:mathcalG} and \eqref{eq:mathcalF}, using the steady-state $ \bar{\nu} $ profile. The results show that the density enhancement originates from a negative radial gradient in viscous torque ($ \partial_R \mathcal{G} < 0 $), which drives an outward-directed mass flux from the outer edge of the active zone into the dead zone. \\
\indent In essence, this outward drift resembles the expansion of the outer edge of a finite-sized viscous accretion disc. The material near the dead--active zone interface receives angular momentum from material interior to it, but due to the lower $\alpha$ in the dead zone, lacks an efficient mechanism to transport that momentum further outward. Consequently, this material itself must move outward. In this way, the dead--active zone interface effectively acts as a barrier to outward angular momentum transport. Furthermore, despite being governed by viscous processes, the short radial length scales, $ \delta R $, involved, mean that the local viscous timescale, $\tau_{\text{visc}}$, can approach the dynamical timescale, since $ \tau_{\text{visc}} \!\sim\! \delta R^2 / \bar{\nu} \!\sim\! \alpha^{-1} (\delta R / H)^2 \Omega_\text{K}^{-1} $. \\
\indent Interestingly, both panels of Fig.~\ref{fig:sigma_evolution_compare_1D_versus_3D} reveal a fixed point where the net radial mass flux is zero, so that $\Sigma$ is constant. This coincides with a turning point in the mass flux profile (red line in Fig. \ref{fig:1D_diffusion_model_F_G_comparison}), highlighting a generic feature of the initial flow structure in this region. \\
\begin{figure}
    \includegraphics[width=0.98\columnwidth]{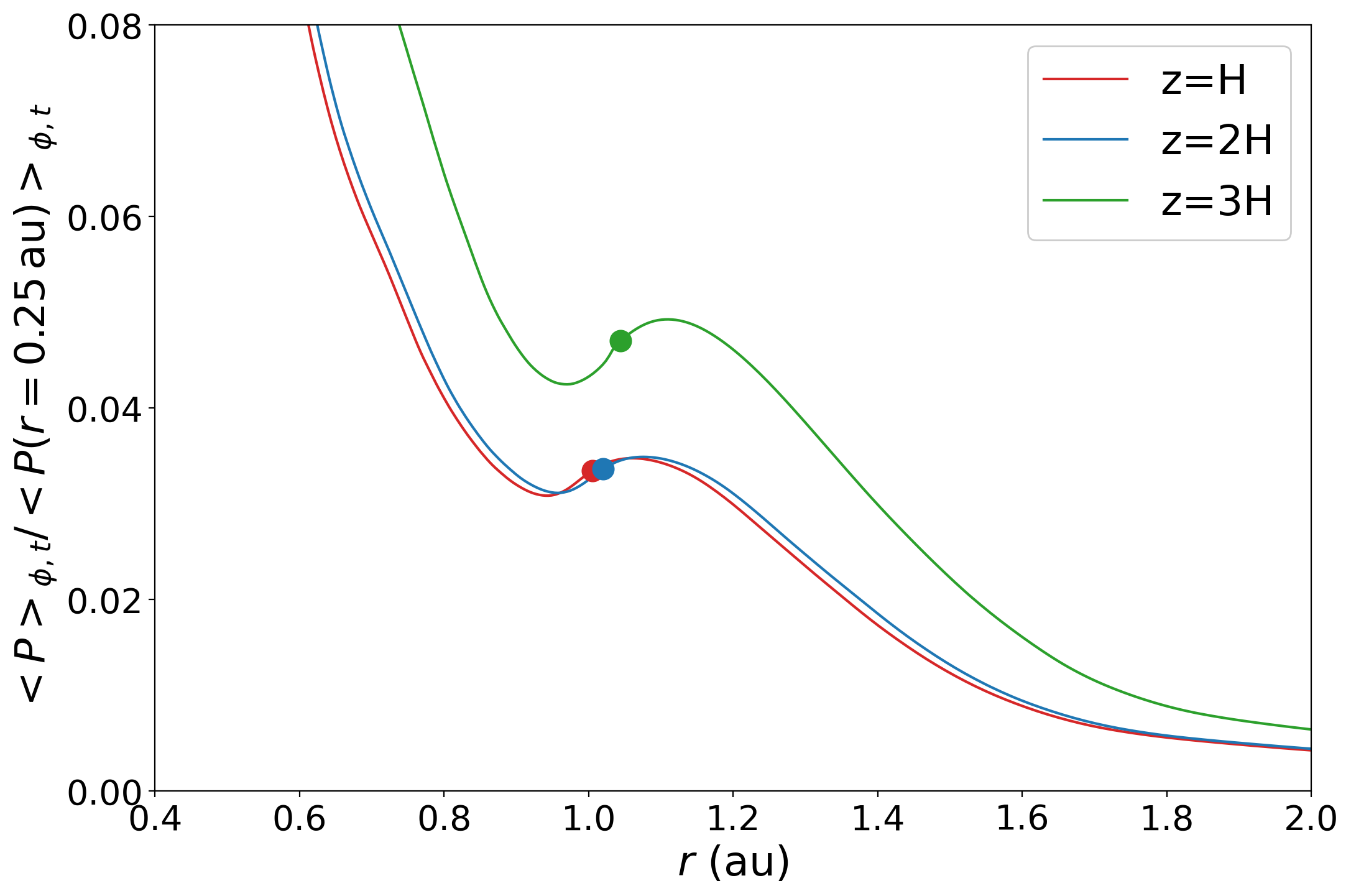}
    \caption{Radial profiles of normalised pressure along the following slices of constant $\theta$ for \texttt{ZNF-FID}: $z=H$ (red), $z=2H$ (blue) and $z=3H$ (green). The local axisymmetric maximum that forms just outside the dead--active zone interface (filled dots) is coherent through $z=\pm3H$, and the temporal average is taken over the interval $t_{\text{in}}\in[600,700]$.}
    \label{fig:pressue_comparison_scale_heights_t600_700}
\end{figure}
\indent The axisymmetric pressure maximum is sustained throughout the duration of the simulations and remains vertically coherent through $z\approx\pm 3.5H$. This is illustrated (up to $z=\pm3H$) by the normalised radial pressure profiles, $\langle P \rangle_{\phi,t}/\langle P (r=0.25\,\text{au})\rangle_{\phi,t}$, along the slices of constant $\theta$ shown in Fig. \ref{fig:pressue_comparison_scale_heights_t600_700}. Notably, this coherence persists despite the complex vertical structure of the MRI-active zone discussed in Section~\ref{section:vertical_accretion_structure}. \\ 
\indent It appears that the sufficiently strong ambipolar diffusion profile (see Fig. \ref{fig:non_ideal_MHD_setup}) suppresses MRI activity in the surface layers of the dead zone, unlike in the Ohmic-resistive three-dimensional simulations of \citet{flock_3d_2017}, thereby enabling the formation of this vertically extended pressure structure. This has important implications for dust dynamics, as it can reduce the radial drift of solids across the vertical extent of the disc at the dead--active zone interface.

\begin{figure*}
    \centering
    \includegraphics[width=1.0\textwidth]{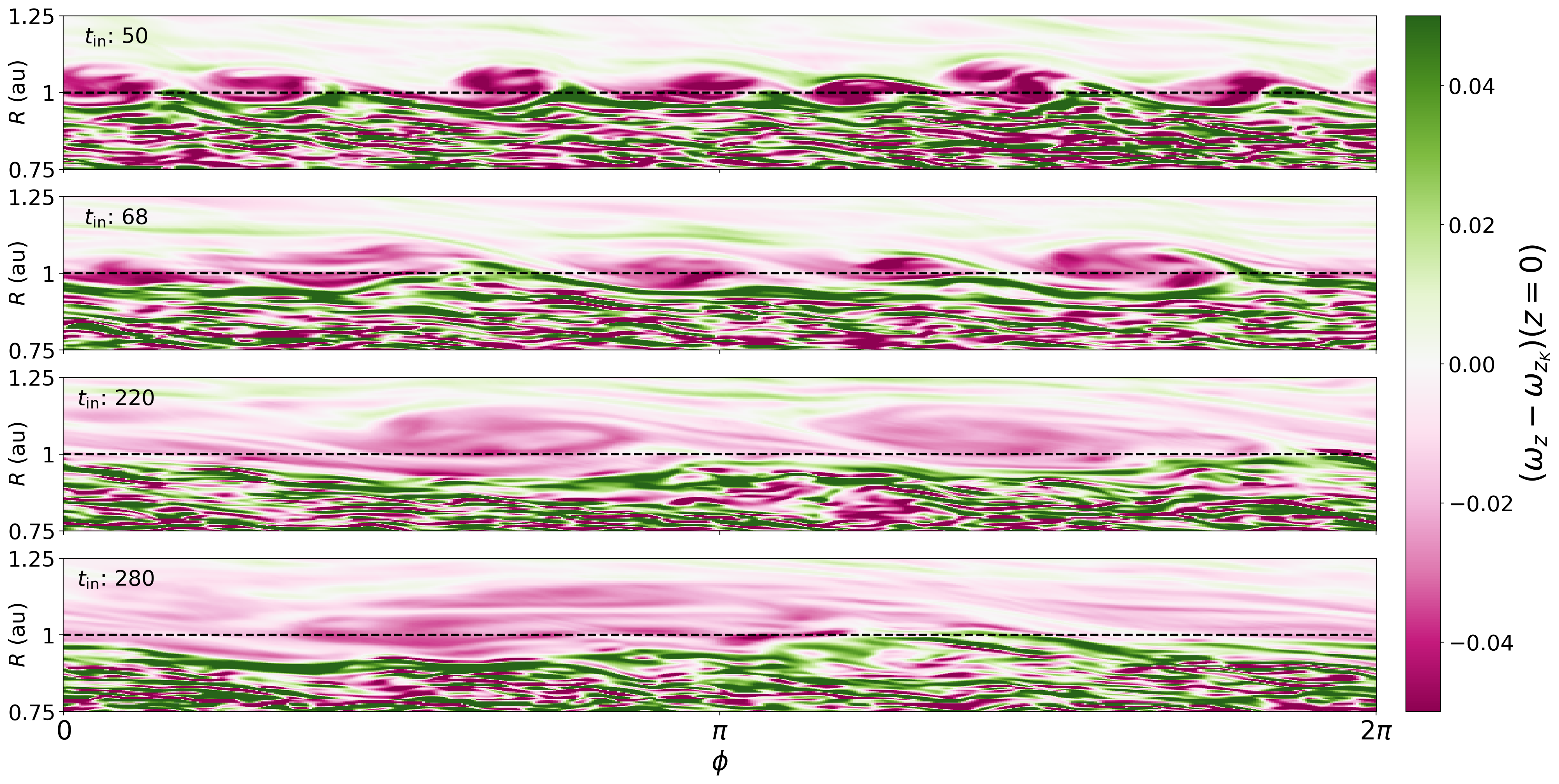}
    \caption{Snapshots of the residual vertical vorticity $(\omega_z-\omega_{z_\text{K}})$ at the disc midplane for \texttt{ZNF-FID}. From top to bottom, the dominant azimuthal mode structures are $m=7$, $m=5$, $m=2$ and $m=1$, in agreement with Fig. \ref{fig:test10_delta_vphi_spectrum_R4.4}. The anticyclonic vortices encroach upon the MRI-active region (the dashed black line denotes the dead--active zone interface) and the eventual $m=1$ vortex possesses a weak and disrupted spatial morphology (bottom panel) during its lifetime.}
    \label{fig:early_life_vortex_midplane_plots}
\end{figure*}
\subsection{Vortices}
\label{section:vortices}

\subsubsection{Formation}
\begin{figure}
    \includegraphics[width=0.98\columnwidth]{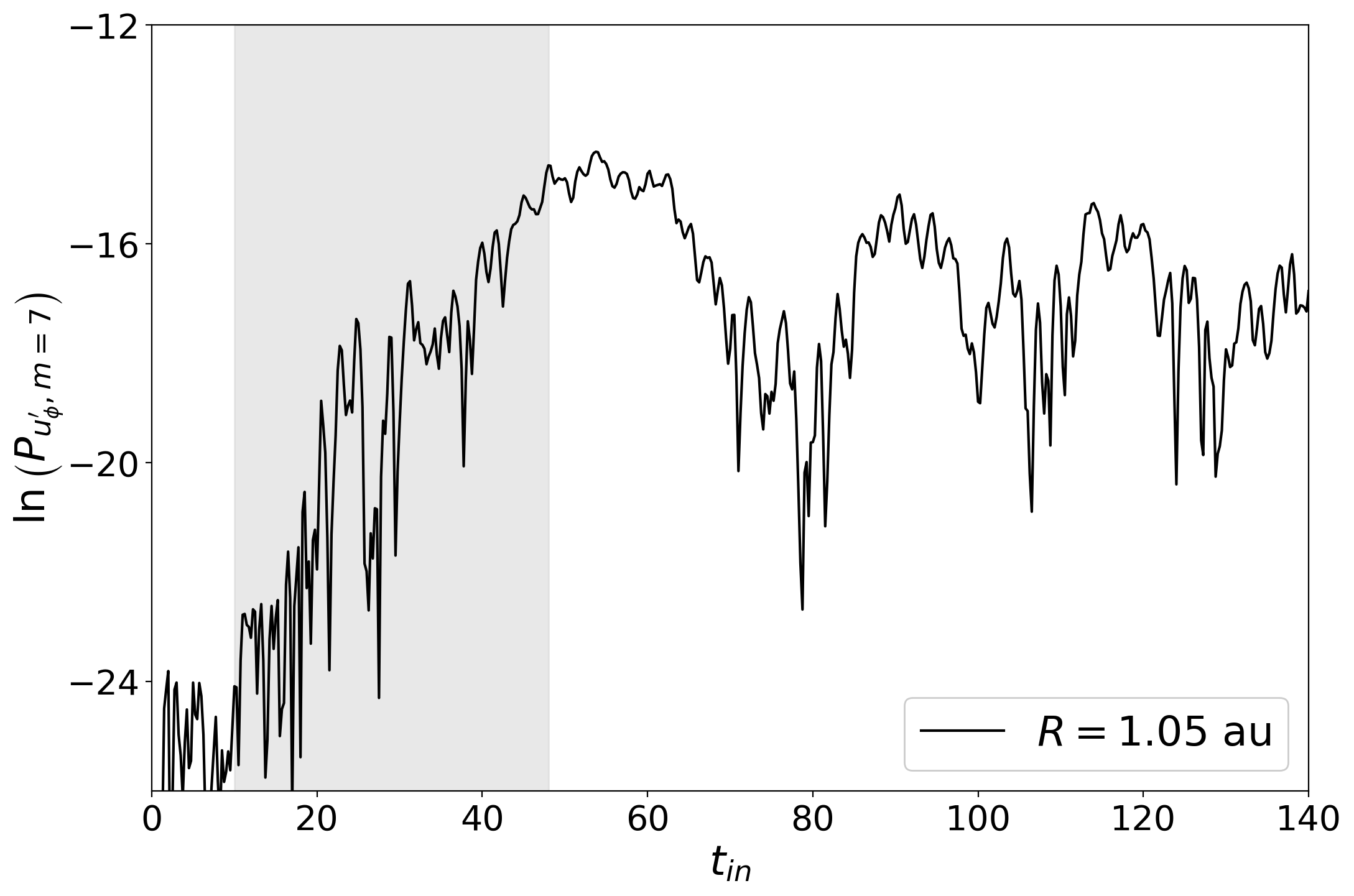}
    \caption{Fourier-power time series of the azimuthal velocity deviation from Keplerian flow $P_{u_\phi',m=7}\,(R = 1.05\,\text{au}, t)$, for the fastest growing Rossby wave mode ($m=7$), evaluated at the vortex centres. The grey-shaded region indicates the linear growth phase of the RWI, with an approximate growth rate of $\sigma\sim0.17\Omega_\text{K}$.}
    \label{fig:m7_vortex_location_growth_rate}
\end{figure}

\begin{figure}
    \includegraphics[width=0.98\columnwidth]{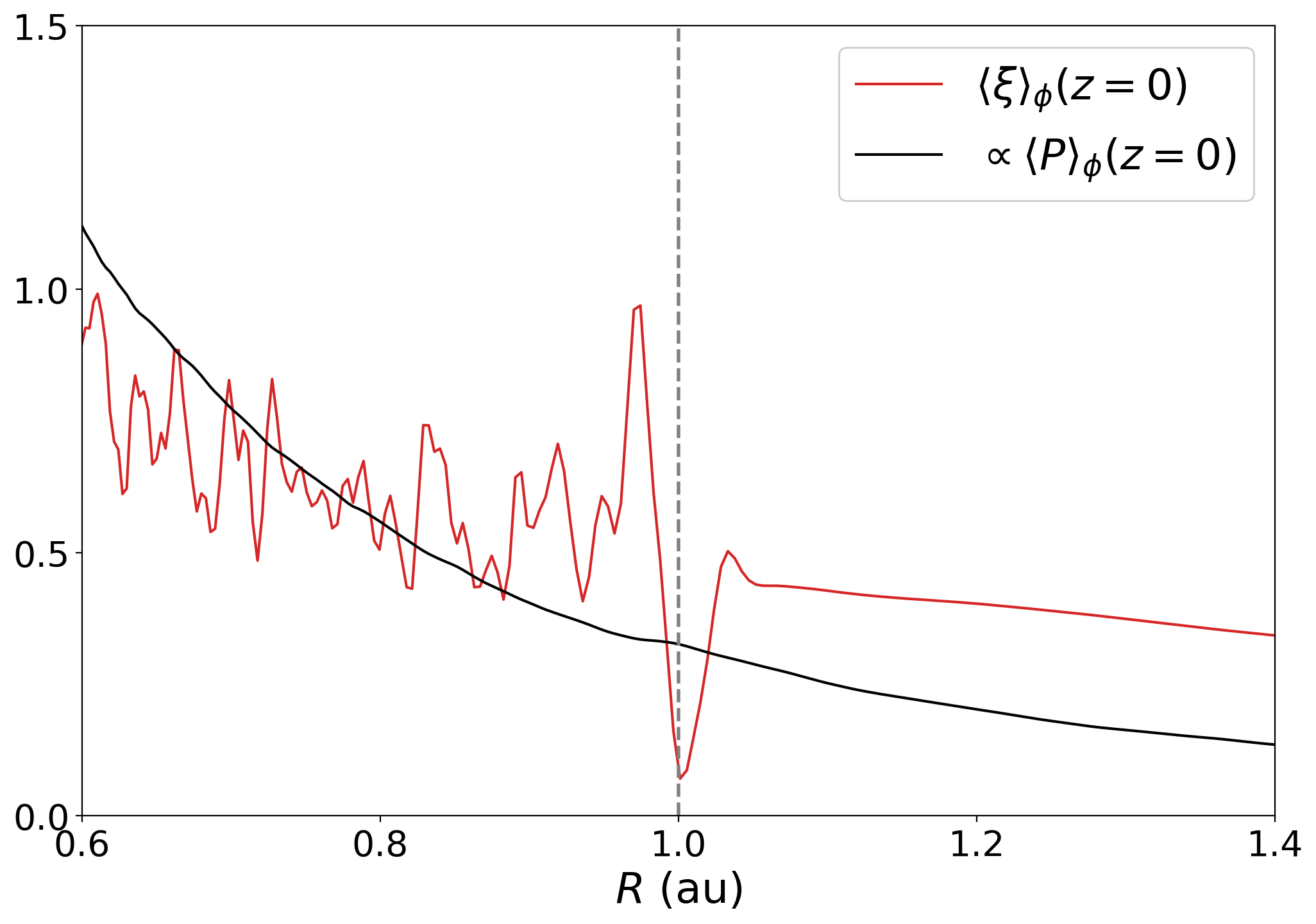}
    \caption{Radial profiles of the azimuthally averaged vortensity, $\langle \xi \rangle_\phi$ (red), and pressure, $\langle P \rangle_\phi$ (black), at the midplane. The profiles are taken at $t_{\text{in}}=10$, which corresponds to the start of the linear growth phase of the RWI (see Fig. \ref{fig:m7_vortex_location_growth_rate}). Close to the dead--active zone interface (dashed grey line) there is a distinct minimum in the vortensity and no local pressure maximum. The pressure profile is normalised by a factor of $10^3$.}
    \label{fig:pressure_vortensity_comparison_tin10_RWI_moment}
\end{figure}

Anticyclonic vortices form at the dead--active zone interface in all four dead--active zone interface simulations. This is illustrated for \texttt{ZNF-FID} in Fig. \ref{fig:early_life_vortex_midplane_plots}, which shows snapshots of the residual vertical vorticity $(\omega_z-\omega_{z_\text{K}})$ at the midplane, where $\omega_{z_\text{K}}=\Omega_\text{K}/2$ is the Keplerian vertical vorticity. \\
\indent These vortices arise from the saturation of the RWI \citep{lovelace_rossby_1999, lovelace_rossby_2014}. The variation of RWI modes in $\phi$ can be captured with an azimuthal Fourier-mode decomposition, characterised by the wave number $m$, as defined in equation~\eqref{eq:fourier_m_mode_decomposition}. In our setup, the $m=7$ mode grows fastest, initially forming seven coherent anticyclonic vortices (see top panel of Fig. \ref{fig:early_life_vortex_midplane_plots}). \\
\indent To compare with linear theory, we examine the Fourier-power time series of the azimuthal velocity deviation, $u_\phi'$, defined in Section~\ref{section:averages_asymmetries_magnetic_fields}, evaluated at the vortex centres: $ P_{u_\phi',m}\,(R = 1.05\,\mathrm{au}, t) $. This is computed across modes with $ m\!<\!10 $, as defined in equation~\eqref{eq:fourier_m_mode_decomposition}. Fig.~\ref{fig:m7_vortex_location_growth_rate} shows the initial power time series for the $m=7$ mode, which undergoes exponential growth followed by saturation at $t_{\text{in}}\!\sim\!50$, consistent with the residual vertical vorticity plots in Fig. \ref{fig:early_life_vortex_midplane_plots}. The growth rate of the $m=7$ mode, corresponding to a perturbation $\propto\exp(\sigma t)$, is $\sigma\!\sim\!0.17\,\Omega_\text{K}(R=1.05\,\text{au})$, which is roughly consistent with \citet{li_rossby_2000}. Additional measurements show slower growth rates for smaller and larger $m$. In contrast with \citet{li_rossby_2001}, we do not find the $m=4$ or $m=5$ modes to be the fastest growing. Instead, the rapid reshaping of the vortensity profile at the interface produces a sharper, narrower minimum, favouring a higher $m=7$ mode. \citet{ono_parametric_2016} showed that dominance of the $m=7$ mode places the system near the Rayleigh stability threshold under their assumptions; however, it remains stable throughout our simulation, despite the sharp interface. \\
\indent Whilst there is no general analytic criterion for the onset of the RWI, it is the presence of a vortensity minimum, and not a pressure maximum that is critical. We confirm this in Fig. \ref{fig:pressure_vortensity_comparison_tin10_RWI_moment}, which shows radial profiles of the pressure, $\langle P\rangle_\phi$, and vortensity, $\langle \xi \rangle_\phi$, defined in equation~\eqref{equation:vortensity_definition}, at $t_{\text{in}}=10$, corresponding to the onset of the RWI's linear phase (see Fig. \ref{fig:m7_vortex_location_growth_rate}). A distinct localised vortensity minimum is present just outside the dead--active zone interface, with minimal variation in the pressure structure. \\
\indent The rapid onset of the RWI is also confirmed in \texttt{ZNF-MRI-IC}, demonstrating that it is not simply a consequence of the initial $\rho$ or $\mathbf{B}$ profiles. Moreover, we find consistency with the criterion proposed by \citet{chang_origin_2023}, which predicts that the RWI is triggered when the radial oscillation frequency -- the epicyclic frequency for a barotropic flow -- is $\lesssim75$ per cent of the local Keplerian frequency. \\

\subsubsection{Initial evolution and vortex morphology}
\begin{figure}
    \includegraphics[width=0.98\columnwidth]{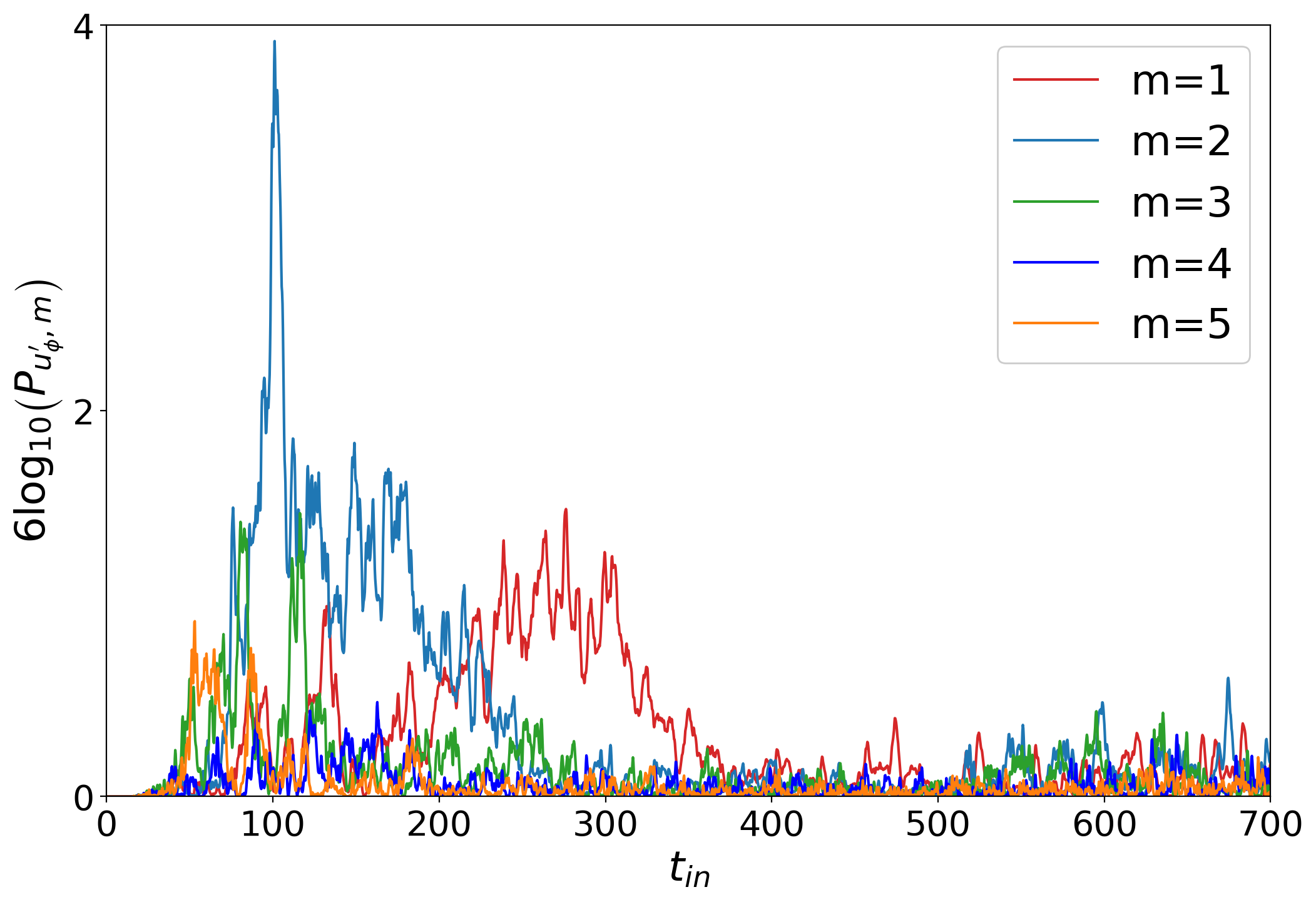}
    \caption{Fourier-power time series of the azimuthal velocity deviation from Keplerian flow, $P_{u_\phi',m}\,(R = 1.05,\text{au}, t)$, evaluated at the vortex centres for azimuthal mode numbers $m \in \{1, 2, 3, 4, 5\}$. Compare the vortex succession process shown here with Fig.~\ref{fig:early_life_vortex_midplane_plots}, and note the subsequent decay of the power budget associated with the weakly coherent 
    $m=1$ structure. The power is normalised by a factor of $10^6$.}
    \label{fig:test10_delta_vphi_spectrum_R4.4}
\end{figure}

Radial profiles of the Fourier-power time series, $P_{u_\phi',m}$, for the azimuthal mode numbers $m \in \{1, 2, 3, 4, 5\}$, are shown in Fig. \ref{fig:test10_delta_vphi_spectrum_R4.4}. Whilst, the $m=7$ mode dominates initially, its contribution is negligible after $t_\text{in}\!\sim\!65$ and is omitted for clarity. The power distribution then sequentially evolves through coherent $m=5$ and $m=2$ phases, consistent with the snapshots shown in Fig. \ref{fig:early_life_vortex_midplane_plots}, before evolving into an $m=1$ structure by $t_\text{in}\!\sim\!230$. This corresponds to a vortex-merger timescale of approximately $20$ local orbits. \\
\indent The $m=1$ anticyclonic vortex possesses a weak and spatially incoherent morphology (see bottom panel of Fig.~\ref{fig:early_life_vortex_midplane_plots}). We attribute this weakened structure to the turbulent dynamics and associated wave excitation within the MRI-active zone \citep[e.g.][]{heinemann_excitation_2009}, into which the nearby vortices encroach. \\
\indent Additionally, the vortices do not migrate and remain fixed at the localised pressure maximum, in agreement with \citet{paardekooper_vortex_2010}. This behaviour is different from the migrating vortices seen in more thermodynamically complex simulations by \citet{faure_vortex_2015}. \\
\indent Finally, through $m$-mode analysis of the disc midplane, we establish a connection between the spiral density waves in the dead zone and the vortices, as at any instant, the waves exhibit the same $m$ as the vortices that launch them. These waves contribute to the midplane $\alpha_\mathcal{R}$ on the order of $10^{-4}$ in the dead zone shown in Fig.~\ref{fig:test10_midplane_averages_800_2800}, offering a weak mechanism for angular momentum transport. They may also aid in the vertical lofting of dust and intermittently concentrate solids within their crests.
\subsubsection{Long-term evolution}
\begin{figure}
    \includegraphics[width=0.98\columnwidth]{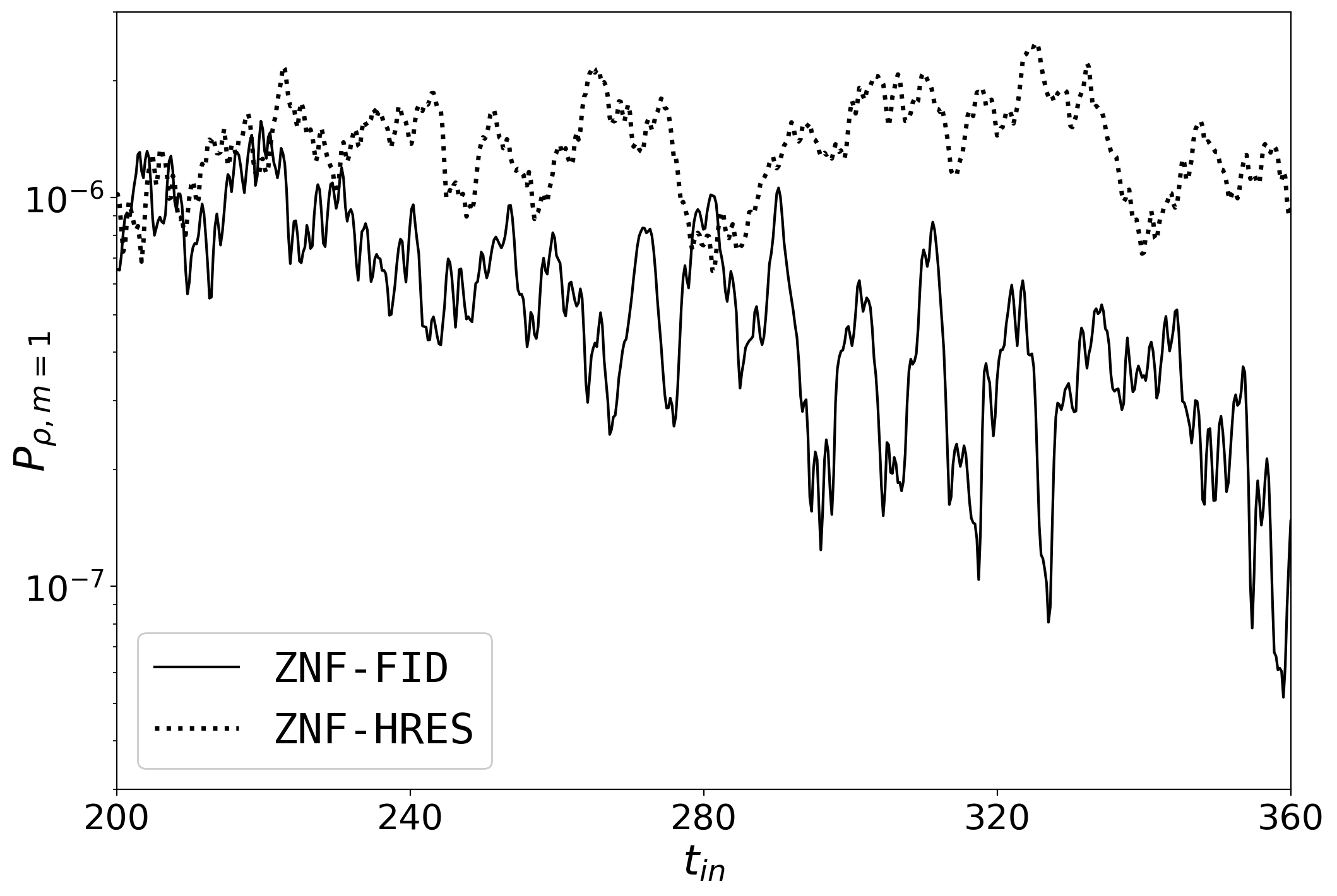}
    \caption{Fourier-power time series of the $m=1$ density mode, $P_{\rho,m=1}\,(R = 1.05,\text{au}, t)$, evaluated at the vortex centres. The power (and $m=1$ vortex) decays more rapidly in \texttt{ZNF-FID} (solid), compared to the higher-resolution test \texttt{ZNF-HRES} (dotted).}
    \label{fig:resolution_power_spectrum_test}
\end{figure}

Following the merger process, the weakly coherent $m=1$ vortex shears out over a timescale of $\sim\!40t_{\text{DZI}}$, with its power budget redistributed into the axisymmetric ($m=0$) mode. This timescale is comparable to that reported by \citet{flock_gaps_2015}, who used the same azimuthal domain extent in their outer dead zone edge study. \\
\indent To evaluate the influence of numerical resolution on the rapidity of this decay, we performed a resolution test (\texttt{ZNF-HRES}), with double the number of cells in azimuth and radius. We then analyse the Fourier-power spectrum of the density field, focusing on the $m=1$ mode, $P_{\rho,m=1}$, defined in Section \ref{section:averages_asymmetries_magnetic_fields}, and present the results in Fig. \ref{fig:resolution_power_spectrum_test}. This reveals an order-of-magnitude difference in the evolution of the $m=1$ power across resolutions, highlighting the sensitivity of vortex persistence to numerical resolution at this level. \\
\indent We also consider how vortex evolution may proceed in the high-resolution limit of global simulations. Multiple vortex stages observed in our simulations are likely susceptible to different variants of the elliptical instability \citep{pierrehumbert_universal_1986, kerswell_elliptical_2002, lesur_stability_2009, mizerski_magnetoelliptic_2009, railton_local_2014}: the three-dimensional $m=7$ vortices exhibit relatively low aspect ratios ($\sim\!3$), whereas the later-stage $m=1$ vortex displays higher aspect ratios ($>\!5$). Whilst we see no evidence of elliptical instability at the resolutions or timescales considered, it may emerge at higher resolutions or over longer integration times \citep[e.g.][]{richard_structure_2013}. If triggered, it could render the vortex cores turbulent, disrupting dust aggregation at their centres and potentially destroying the coherent vortical streamlines. \\
\indent To investigate the potential for vortex regeneration, we extend the runtime in \texttt{ZNF-LONG} to $225\,t_{\text{DZI}}$. Whilst renewed vortensity activity appeared at the dead--active zone interface, following the dissipation of the initial vortex, no subsequent development of a vortensity structure capable of triggering the RWI occurred. A likely explanation is that the relatively small active zone allows the inner boundary condition to interfere with the long-term evolution and radial redistribution of vortensity, both essential for re-establishing the conditions necessary to trigger the RWI. Radially extended simulations, such as the outer dead zone study by \citet{flock_gaps_2015}, show evidence for non-migrating periodic vortex life cycles.

\begin{figure*}
    \includegraphics[width=1.0\textwidth]{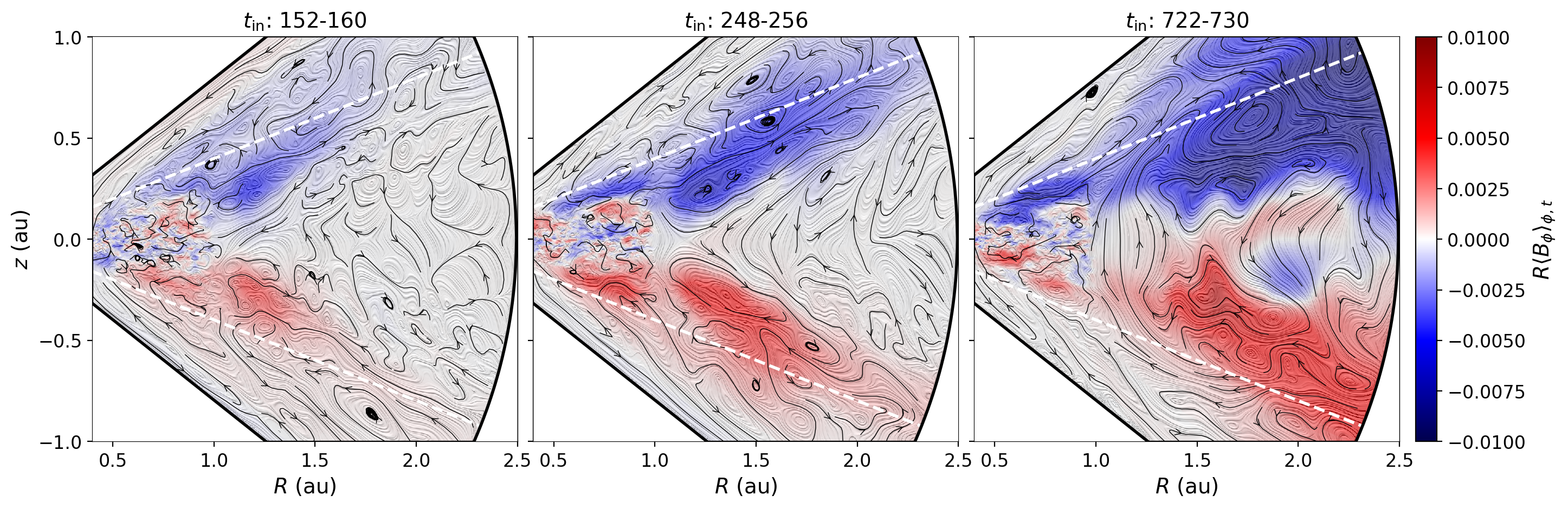}
    \caption{Magnetic field evolution for \texttt{ZNF-FID}, averaged in azimuth and over a single orbit at the dead--active zone interface, as indicated above each plot. The background colour shows the normalised toroidal magnetic field, $R\langle B_\phi \rangle_{\phi,t}$, and the dashed white lines mark the disc--corona transition. The poloidal magnetic field, $\langle\mathbf{B_p}\rangle_{\phi,t}$, shown using black field lines and a LIC overlay in greyscale, reveals coherent large-scale clockwise loops in the dead zone and MRI-active region (see Fig. \ref{fig:BpLIC_BPhi_small_600700}). Note the broad anti-correlation between the sign of the toroidal and radial magnetic fields, and that the disordered poloidal field in the dead zone in the left panel is extremely weak.}
    \label{fig:BpLIC_combined_final_masked}
\end{figure*}
\section{Results III. Magnetic Field Structures in the dead zone}
\label{section:magnetic_field_structures}
We complete the presentation of our results by outlining the evolution of magnetic field structures in the dead zone in Section~\ref{section:magnetic_fields_dead_zone_overview}. In Section~\ref{section:magnetic_fields_poloidal_field} we examine the coherent poloidal magnetic field structures in the dead zone, and their capacity to drive a weak magnetic-pressure wind. Finally, in Section~\ref{section:magnetic_fields_dead_zone_formation_mechanism} we speculate on the formation mechanism of these magnetic field structures in the dead zone. \\
\indent Throughout this section, the analysis focuses on $\texttt{ZNF-FID}$, but we note that the three other dead--active zone interface simulations exhibit a similar permeation of magnetic field structures through the extent of the dead zone.
\subsection{Overview}
\label{section:magnetic_fields_dead_zone_overview}
We begin by briefly examining the global evolution of the magnetic field in the dead zone in \texttt{ZNF-FID}. An initially toroidal magnetic field confined to the active zone (see left panel of Fig. \ref{fig:bx3_evolution}), expands and evolves into a complex, large-scale field structure with both toroidal and poloidal components, which gradually fills the entire dead zone. \\
\indent To capture this evolution in more detail, Fig. \ref{fig:BpLIC_combined_final_masked} presents three snapshots of the field evolution, averaged azimuthally and temporally over a single orbit at the dead--active zone interface. Whilst Fig. \ref{fig:bx3_evolution} only illustrates the spreading of the toroidal field, the addition of poloidal magnetic field lines, $\langle\mathbf{B_p}\rangle_{\phi,t}$, and their associated LIC overlay, in Fig. \ref{fig:BpLIC_combined_final_masked} provides a complementary perspective, highlighting the emergence and organisation of the poloidal component. Please note that the apparently disordered poloidal field present in the dead zone at early times (left panel of Fig. \ref{fig:BpLIC_combined_final_masked}), is extremely weak and an unfortunate result of the visualisation technique. \\
\indent Notably, the growth of $B_\phi$ in the dead zone coincides with the emergence of increasingly ordered poloidal field structures. In particular, we observe a tendency towards anti-correlation between the signs of $B_r$ and $B_\phi$, along with the formation of concentric, coherent large-scale poloidal loops that gradually diffuse outward over time and encircle an \emph{inversion point} (see middle panel of Fig.~\ref{fig:BpLIC_combined_final_masked}). These features are examined in greater detail in the following two sections.

\subsection{Poloidal magnetic field structure in the dead zone}
\label{section:magnetic_fields_poloidal_field}

\subsubsection{Morphology and evolution}
\label{section:magnetic_fields_poloidal_field_dead_zone_morphology_and_evolution}

\begin{figure}
    \centering  
    \includegraphics[width=0.98\columnwidth]{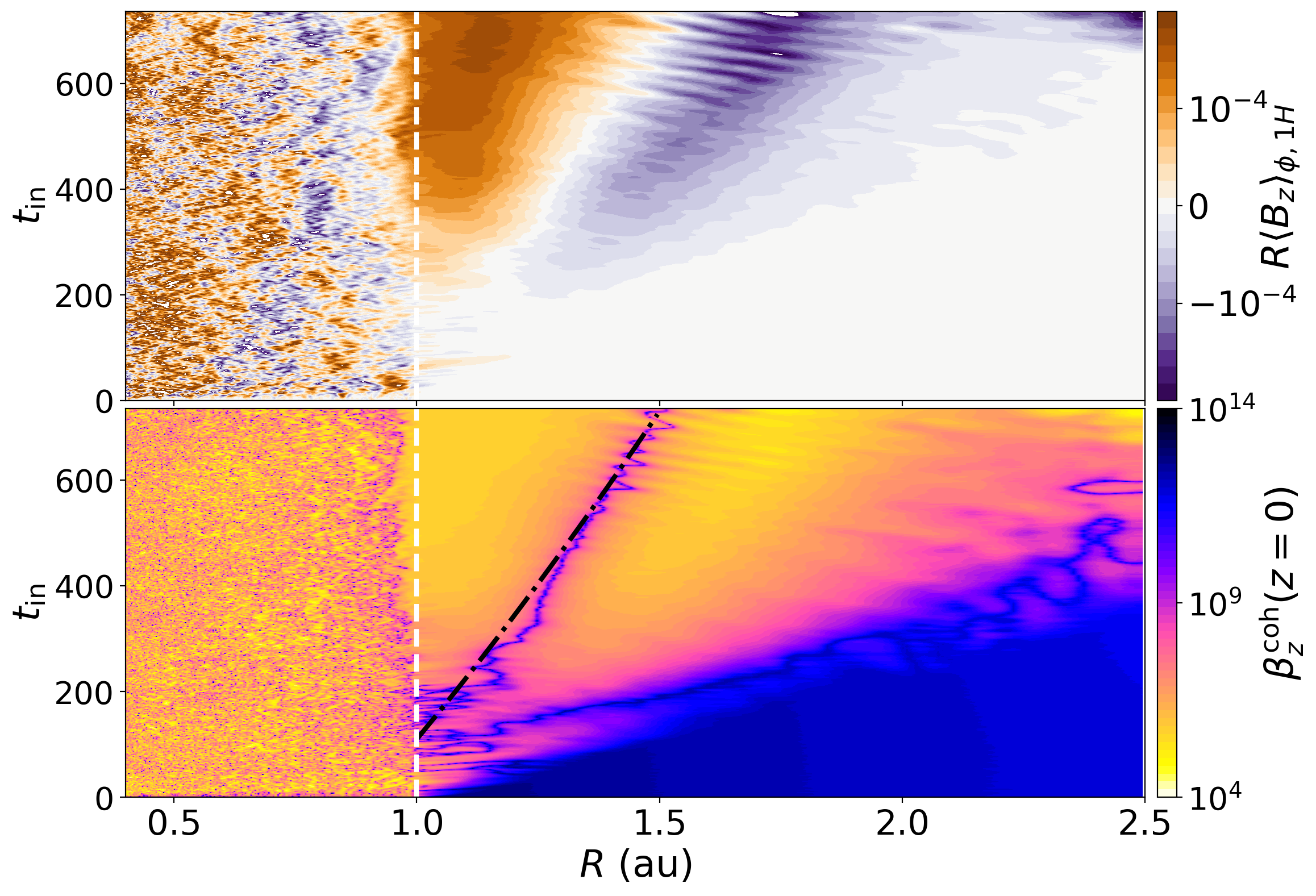} \\
    \caption{Spacetime $(R,t)$ diagrams of the vertical magnetic field evolution. Top: Azimuthally and meridionally integrated (normalised) vertical magnetic field, $R\langle B_z \rangle_{\phi, 1H}$. The trend of upward-oriented field (orange) in the inner active/dead zone and downward-oriented field (blue) in the outer active/dead zone is consistent with the large-scale poloidal field loops of the same orientation in the MRI-active (see Fig.~\ref{fig:BpLIC_BPhi_small_600700}) and MRI-dead (see Fig.~\ref{fig:BpLIC_combined_final_masked}) zones. Bottom: The coherent vertical component of the plasma-$\beta$ parameter, $\beta_z^{\mathrm{coh}}$, at the midplane reveals the outward migration of field structures from the dead--active zone interface (dashed white line). The inversion point is well matched by an outward propagation speed of $ 5 \times 10^{-4} u_\text{K} $ (dot-dashed black line), based on a visual fit.}
    \label{fig:combined_Bz_midplane}
\end{figure}

\indent The initially absent poloidal magnetic field in the dead zone gradually develops into a striking and complex configuration by $t_{\mathrm{in}}\!\sim\!250$, as shown in the middle panel of Fig.~\ref{fig:BpLIC_combined_final_masked}. This consists of large-scale, clockwise poloidal loops encircling an inversion point, with ordered vertical field oriented positively on the interior and negatively on the exterior. Throughout the rest of the simulation, this configuration migrates outward, and becomes increasingly coherent: by $t_{\text{in}}\!\sim\!730$, the large-scale clockwise poloidal field loops are clearly identifiable in the dead zone. \\ 
\indent In Fig.~\ref{fig:combined_Bz_midplane}, $(R,t)$ spacetime plots are presented to characterise the radial structure of the vertical magnetic field in the disc. The top panel shows $R\langle{B_z}\rangle_{\phi,1H}$, the vertical field averaged in azimuth and over the meridional direction within a single scale height, $H$, either side of the midplane. This confirms the emergence of an ordered vertical field geometry that extends around the midplane, forming coherent loops in both the active zone (see Fig. \ref{fig:BpLIC_BPhi_small_600700}) and the dead zone, where the loops retain the same orientation, as discussed above. \\
\indent This ordered vertical field in the dead zone migrates outward, with the inversion point developing near the dead--active zone interface, as evident in both panels of Fig.~\ref{fig:combined_Bz_midplane}. The bottom panel shows $\beta_z^{\text{coh}}(z=0)$, the coherent vertical component of the plasma-$\beta$ parameter, as defined in equation~\eqref{eq:beta_vertical_coherent}. The coherent field structure, which we characterise by the inversion point, is transported radially outward at a velocity of roughly $ 5 \times 10^{-4} u_\text{K} $ at the midplane, based on a visual match of its trajectory. This value is broadly consistent with a characteristic Ohmic-diffusion speed, $v_O\sim \eta_\text{O}/R$, yielding $v_\text{O}\!\sim \!(H/R)^2\,\text{R}_\text{m}^{-1}\,u_\text{K}\! \sim \!10^{-4}u_\text{K}$. Also see top-right panel of fig.~11 in \citet{lesur_systematic_2021} and note that their velocities are normalised by an extra factor of $\varepsilon^{-1}$. This suggests that the midplane outward migration is driven by non-ideal MHD diffusion, dominated by the Ohmic component.

\subsubsection{Outflow characteristics in the inner dead zone}
\begin{figure}
    \includegraphics[width=0.98\columnwidth]{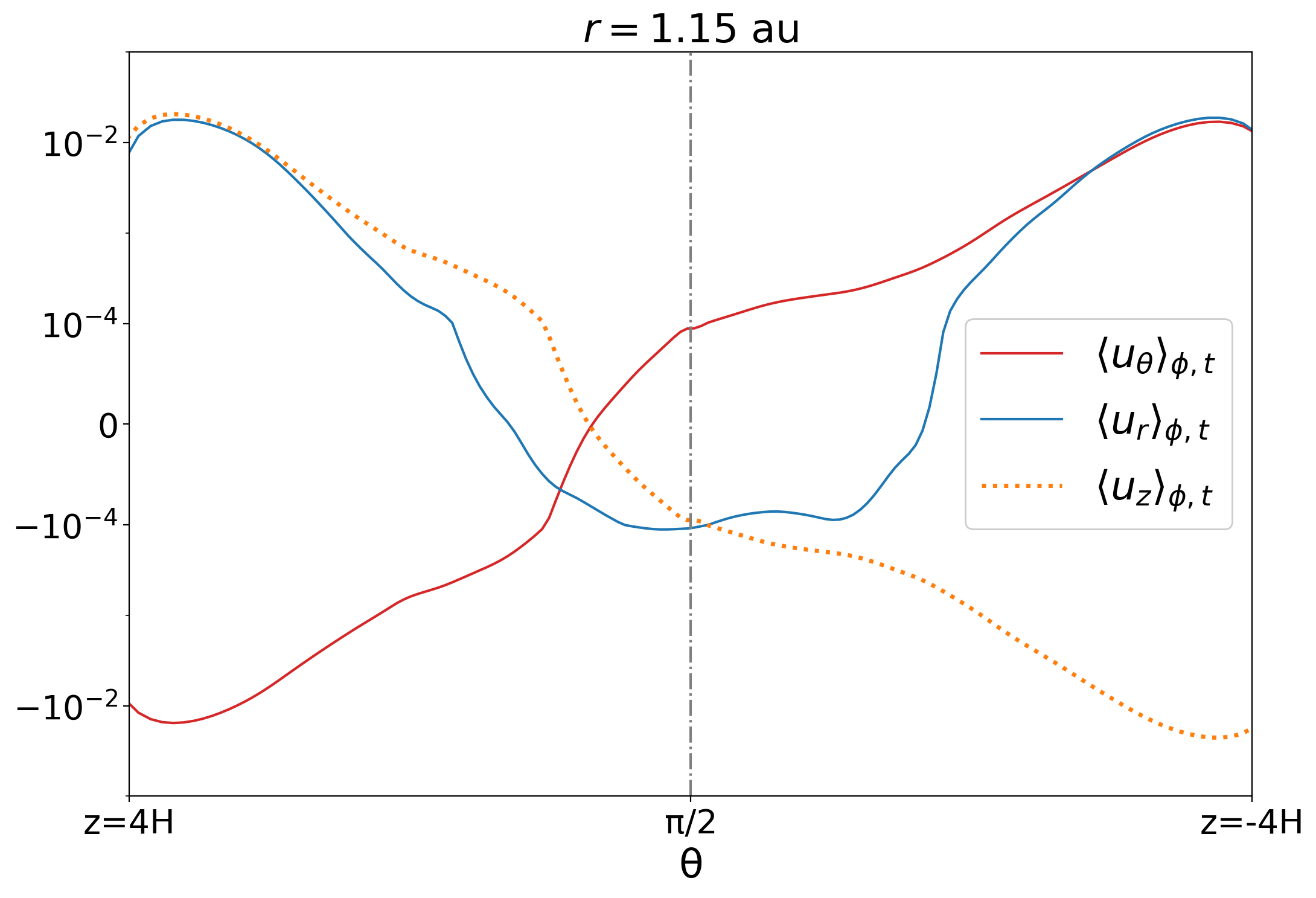}
    \caption{Decomposition of the poloidal velocity field, $\langle\mathbf{u_p}\rangle_{\phi,t}$, into its coordinate components along the spherical contour $r=1.15\,\text{au}$. This contour corresponds with the coherent vertical magnetic field configuration at the inner edge of the dead zone presented in Fig.~\ref{fig:BpLIC_combined_final_masked}, and the components have been averaged in azimuth and over the interval $t_\mathrm{in}\in[600,700]$. The vertical flow (dotted orange line) is directed away from the midplane, which is driven by the magnetic pressure configuration presented in Fig. \ref{fig:magnetic_field_components_squared_log_600_700_1.15au}.}
    \label{fig:velocity_field_components_log_600_700_1.15au}
\end{figure}

\begin{figure}
    \includegraphics[width=0.98\columnwidth]{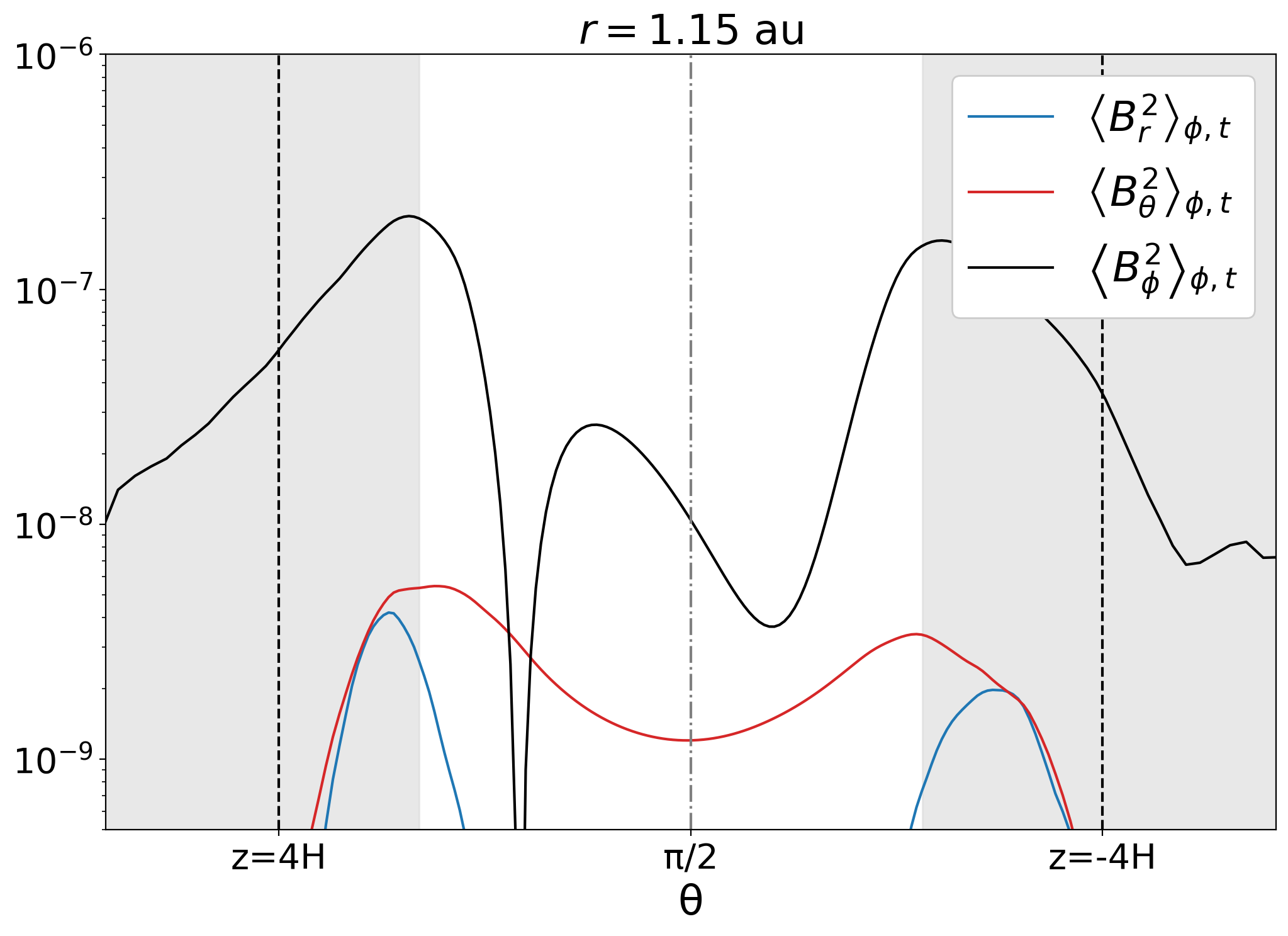}
    \caption{Decomposition of the magnetic energy, $\langle B^2\rangle_{\phi,t}$, into its coordinate components along the spherical contour $r=1.15\,\text{au}$. This contour corresponds with the coherent vertical field configuration at the inner edge of the dead zone presented in Fig.~\ref{fig:BpLIC_combined_final_masked}, and the components have been averaged in azimuth and over the interval $t_\mathrm{in}\in[600,700]$. The grey-shaded regions indicate where the vertical magnetic pressure gradient (dominated by $B_\phi^2$) accelerates the flow away from the disc.}
    \label{fig:magnetic_field_components_squared_log_600_700_1.15au}
\end{figure}

As this complex poloidal field structure expands outward via non-ideal MHD diffusion, the region of the MRI-dead disc threaded by a coherent, vertically oriented field increases (see top panel of Fig.~\ref{fig:combined_Bz_midplane}). By $ t_\mathrm{in}\!\sim\!700 $, the coherent, positively oriented vertical magnetic field extends across the full vertical extent of the disc and spans the radial range, $ 1\,\mathrm{au} \lesssim r \lesssim 1.4\,\mathrm{au} $, as shown in the right panel of Fig.~\ref{fig:BpLIC_combined_final_masked}. The magnetic field topology in this region closely resembles that of vertical-net flux, global, non-ideal MHD simulations (e.g. fig. 2 in \citealt{gressel_global_2020}), which impose a strong vertical laminar torque on the disc, sufficient to drive accretion through the dead zone. In the remainder of this section, we characterise the magnetic field structure along the spherical contour $r=1.15\,\text{au}$, and evaluate its ability to impose vertical torques and drive accretion, in analogy with vertical-net flux configurations. \\
\indent The relative strength of the vertical magnetic field is quantified using the coherent vertical component of the plasma-$\beta$ parameter, defined in equation \eqref{eq:beta_vertical_coherent}. At $r=1.15\,\text{au}$, $\beta_z^\text{coh}\!\sim\!5\times10^5$ for the configuration shown in the right panel of Fig. \ref{fig:BpLIC_combined_final_masked}, indicating a weak but dynamically relevant vertical field. Indeed, the axisymmetric non-ideal MHD simulations of \citet{lesur_systematic_2021} conclude that such $\beta_z^\text{coh}$ values can drive weak magnetic-pressure winds. \\ 
\indent To assess whether this magnetic field configuration drives a magnetic outflow, we first examine the poloidal velocity, $\langle \mathbf{u_p} \rangle _{\phi,t}$, and magnetic energy, $\langle B^2 \rangle _{\phi,t}$, over the spherical contour $r=1.15\,\mathrm{au}$, averaged in azimuth and over the interval $t_\mathrm{in}\in[600,700]$. As shown in Fig.~\ref{fig:velocity_field_components_log_600_700_1.15au}, the flow is directed vertically away from the midplane in the regions where $|z|\gtrsim1H$. Just below the disc's surface layers, the poloidal velocity is $u_\mathrm{p}\!\sim\!2\times{10}^{-2}$, equivalent to the local sound speed in the disc, $c_\text{s}=\varepsilon R^{-1/2}$. Meanwhile, Fig.~\ref{fig:magnetic_field_components_squared_log_600_700_1.15au} shows that the toroidal component of the magnetic energy, $\langle B_\phi^2\rangle_{\phi,t}$, dominates over the other components, reaching a maximum on both sides of the disc midplane, whilst decreasing sharply in the regions $|z| \gtrsim 2.5H$. This configuration supports vertical acceleration by laminar magnetic pressure and identifies the regions around $|z| \approx 2.5H$ as where the weak magnetic field extracts and stores angular momentum. Note that the diffuse off-midplane current sheet associated with this structure (shown in the right panel of Fig.~\ref{fig:BpLIC_combined_final_masked}) imprints a vertical asymmetry on the velocity and magnetic field profiles in Fig.~\ref{fig:velocity_field_components_log_600_700_1.15au}. \\
\indent Furthermore, the vertical flow in the inner dead zone is accompanied by a comparatively weak midplane inward radial flow on the order of $\sim\!0.01c_\text{s}$, consistent with the midplane plots presented in Fig. \ref{fig:test10_midplane_averages_800_2800}. To assess the magnetic contribution to the angular-momentum extraction driving this accretion flow, we quantify the coherent magnetic stresses, defined in equation~\eqref{eq:maxwell_stress_components}, along the contour $r=1.15$ au. At $z=\pm 3H$, the coherent components of the (normalised) Maxwell stress are, $\mathcal{M}^{\text{coh}}_{\theta\phi}/\langle P \rangle_{\phi,t}\!\sim\!\mathcal{M}^{\text{coh}}_{r\phi}/\langle P \rangle_{\phi,t}\!\sim \!2\times10^{-3}$. An additional source of radial angular momentum transport arises from correlated velocity fluctuations near the disc midplane, yielding $\mathcal{R}_{r\phi}/\langle P \rangle_{\phi,t}\!\sim\! 5\times10^{-4}$, as discussed in Section \ref{section:midplane_accretion_structure}. The contribution of each stress component to the mass accretion rate is discussed in \citet{lesur_magnetohydrodynamics_2021}, but is unable to be applied exactly to this evolving, radially restricted region. Nonetheless, within the thin-disc approximation, vertical stresses contribute $\sim\!R/H$ more to the accretion rate than vertically integrated radial stresses. Therefore, coherent vertical angular-momentum extraction is a non-negligible driver of accretion at this radius. \\
\indent In summary, vertical acceleration of the flow in this region is due to a weak magnetothermal wind \citep[e.g.][]{bai_magneto-thermal_2016}; driven by a combination of the vertical laminar magnetic pressure gradient, $\partial_z (B^2) /8\pi$, which is dominated by $B_\phi$, and the thermal pressure acceleration provided by the hot corona, closer to the disc surfaces.

\subsection{Formation mechanism}
\label{section:magnetic_fields_dead_zone_formation_mechanism}
We now speculate on the formation mechanism of the large-scale magnetic field structures in the dead zone, as examined in the preceding parts of this section. We distinguish between different proposed formation pathways, dividing our analysis into the development of the poloidal and toroidal components separately. \\
\indent First, we focus on the development of poloidal magnetic fields in the dead zone. The meridional plots of $\langle{\mathbf{B_p}}\rangle_{\phi,t}$ shown in Fig.~\ref{fig:BpLIC_combined_final_masked}, reveal that the coherent, ordered poloidal field originates in the inner dead zone (see left panel) and gradually diffuses outward. The resulting large-scale coherent field loops possess the same orientation (clockwise) as those in the active zone (see Fig.~\ref{fig:BpLIC_BPhi_small_600700}). Meanwhile, the ($R,t$) spacetime diagram of $\beta_z^{\text{coh}}$, shown in the bottom panel of Fig.~\ref{fig:combined_Bz_midplane}, provides further evidence for this origin: both the inversion point and the vertical field threading the disc midplane trace back to the region closest to the dead--active zone interface. Therefore, based on this evidence, we hypothesise that the large-scale poloidal field configuration in the dead zone originates from coherent field loops in the active zone, as outlined below. \\
\indent Using Fig. \ref{fig:BpLIC_combined_final_masked} as a reference, we propose the following sequence. Turbulent diffusion in the active zone transports ordered poloidal magnetic fields outward across the dead--active zone interface. Beyond this transition, non-ideal MHD diffusion controls the transport of these fields. The outward transport is stronger in the disc surface layers because the diffusion coefficients at the disc surface (dominated by $\eta_\text{A}$) are greater than at the midplane (dominated by $\eta_\text{O}$): $\eta_\text{A}(z=4H)/\eta_\text{O}(z=0)\sim \text{R}_\text{m}(z=0)/\beta(z=4H) > 1$. Furthermore, the field lines in the surface layers are more bent, leading to shorter local length scales, and faster diffusive speeds. Together, these effects produce a \emph{bow-like} structure, with field lines curving back towards the midplane near the dead--active zone interface (left panel). The inversion point subsequently develops at the midplane, likely due to reconnection (middle panel). The resulting complex field structure then expands and gradually diffuses outward (see Section~\ref{section:magnetic_fields_poloidal_field_dead_zone_morphology_and_evolution}), ultimately forming the concentric large-scale poloidal field loops (right panel).  \\
\indent Second, we examine the development of toroidal magnetic fields in the dead zone. The meridional plots of $R\langle{B_\phi}\rangle_{\phi,t}$ in Fig. \ref{fig:BpLIC_BPhi_small_600700} and Fig. \ref{fig:BpLIC_combined_final_masked} show that $B_\phi$ gradually emerges within the dead zone, and that its generation is broadly anti-correlated with the sign of $B_r$. These trends suggest that the toroidal field is primarily generated by the $\Omega$-effect acting on $B_r$, rather than through the outward diffusion of toroidal flux from the active zone. The characteristic timescale for winding by the $\Omega$-effect ($\sim \!\Omega_\text{K}^{-1}$) is shorter than the Ohmic-diffusion timescale, as the magnetic Reynolds number is relatively large ($\text{R}_\text{m} \gg 1$), with the initial $\text{R}_\text{m}$ profile shown in Fig.~\ref{fig:non_ideal_MHD_setup}. Consequently, the generation of $B_\phi$ can be roughly approximated by $B_\phi(t) \sim -B_r \Omega t$. In the outer midplane region of the disc, where magnetic diffusion is weakest and beyond the inversion point, the toroidal field develops over the interval $t_{\text{in}} \in [200,700]$ (see Fig. \ref{fig:bx3_evolution}). During this period, $B_r$ remains approximately steady at $\sim\!5 \times 10^{-5}$, leading to an expected $B_\phi$ amplitude of $\sim\!10^{-3}$ after $\sim\!25$ local orbits, consistent with the values measured from the simulation. For toroidal fields that grow over longer timescales (see the dead-zone surface layers in Fig.~\ref{fig:BpLIC_combined_final_masked}), amplification is ultimately mediated by ambipolar diffusion, which scales with $B^2$. \\
\indent In summary, the formation of large-scale poloidal and toroidal magnetic structures in the dead zone appears to result from distinct mechanisms: turbulent and non-ideal MHD diffusion drives outward transport of poloidal fields from the inner disc, whilst differential rotation acts on radial magnetic fields (the $\Omega$-effect).\\
\indent Finally, whilst we have proposed a mechanism for the development of these large-scale magnetic field structures in the dead zone, this intriguing phenomenon warrants further investigation in more controlled settings -- specifically, simulations with meridional domains closed at the poles and different initial magnetic field configurations. 

\section{Conclusion}
\label{section:conclusion}
We performed five three-dimensional global ZNF non-ideal MHD simulations, to investigate the dynamics around the dead--active zone interface. These simulations used physically motivated prescriptions for Ohmic and ambipolar diffusion, as well as a disc--corona temperature transition. Our analysis focused on three key aspects: the structure of the inner disc, the formation and morphology of large-scale hydrodynamic features at the dead--active zone interface, and the evolution of large-scale coherent magnetic fields in the dead zone. Below, we summarise the main results:

\begin{enumerate}[align=left, labelindent=0pt, itemsep=5pt]
    \item The ZNF MRI dynamo generates coherent, large-scale poloidal field loops that expand to fill the entire MRI-active region, supporting the phenomenology proposed by \citet{jacquemin-ide_magnetorotational_2024}. Our results also highlight the potential role of the dead zone in mediating the outward propagation of this generated field, a regime not explored in previous studies.
    \item These large-scale poloidal loops are vertically bounded by the disc--corona temperature transition in the MRI-active region, leading to the accumulation of strong, tightly wound magnetic fields in the surface layers. This modifies the accretion profile in the active region, producing a significant component of surface-layer accretion. We also demonstrate that this behaviour persists in fully MRI-active discs, suggesting that the vertical buoyancy configuration imposed by a corona transition plays a broader role in shaping disc dynamics. Finally, we speculate that incorporating a sufficiently heated corona could significantly influence the elevated disc structures seen in global vertical-net flux MRI simulations \citep{zhu_global_2018, jacquemin-ide_magnetic_2021}. 
    \item The formation of an axisymmetric pressure maximum that extends through multiple disc scale heights is driven by a localised density enhancement, resulting from sustained outward mass flux at the outer edge of the MRI-active region. 
    \item Rossby-wave-induced vortices consistently form at the dead--active zone interface in the ZNF regime. They encroach upon the active region and are only weakly coherent due to turbulent disruption at the interface. These vortices eventually decay, primarily through numerical dissipation. The robustness of the formation mechanism for both large-scale HD structures in the presence of an initial vertical-net flux will be investigated in the second paper of this series. 
    \item The ZNF toroidal magnetic field, initially confined to the active region, develops a complex morphology that gradually permeates the entire dead zone. Notably, a coherent large-scale vertical magnetic field emerges, threading the inner dead zone of the disc. Although the formation mechanism and robustness of this configuration remain uncertain, this represents the first instance of a self-consistently generated magnetic structure capable of launching a weak magnetic-pressure wind from the dead zone in protoplanetary disc simulations. 
\end{enumerate}

\indent Global simulations provide valuable insight, but come with important caveats. They isolate a section of the disc from its broader environment, evolve it from idealised initial conditions, and typically span only a short timescale relative to the disc's secular evolution. Moreover, their ability to systematically explore the parameter space -- such as boundary conditions, domain size, temperature prescriptions, and initial magnetic configurations -- is inherently limited. These issues are particularly relevant in the inner disc, where our thermodynamically simple results must be considered in the context of the observed variability, and the connection to the complex disc--star boundary \citep[e.g.][]{zhu_global_2025, takasao_connecting_2025}. \\
\indent Whilst we have proposed mechanisms for the development of large-scale magnetic field structures in our simulations -- including surface-layer accretion near the disc--corona transition in the active zone, and the permeation of the dead zone by magnetic fields -- both of these intriguing phenomena merit further investigation in more controlled settings. In particular, 
the generation of large-scale poloidal field loops within the active region at smaller disc aspect ratios, and the influence of a hot corona on the MRI-driven dynamo, both warrant further study. \\
\indent In addition to these outstanding questions, our results suggest several promising directions for future work within both the ZNF and vertical-net flux regimes. Key extensions include incorporating the Hall effect, due to its known dynamical influence in the dead zone \citep{wurster_we_2021}; implementing a dust--gas feedback module to explore dust aggregation in the large-scale HD structures; and introducing a mobile dead--active zone interface \citep[e.g.][]{faure_vortex_2015}, to probe the interpenetration of dynamics and thermodynamics in this region. \\
\indent However, our immediate next step is to extend the present study to the vertical-net flux regime in the second paper within this series, similar to that recently explored by \citet{iwasaki_dynamics_2024}. There, we focus on global magnetic flux transport, strong variability at the dead--active zone interface, and the interplay between laminar and turbulent flows in the inner disc -- particularly in relation to outflows, accretion, and the formation of large-scale hydrodynamic structures. \\
\indent Together, these papers build a more complete picture of how magnetically mediated processes around the dead--active zone interface shape the structure and evolution of the inner regions of protoplanetary discs.


\section*{Acknowledgements}
The authors would like to thank the anonymous referee for useful comments that helped improve the manuscript.\\
\indent This research was supported by a PhD studentship from the Science and Technology Facilities Council (STFC), grant number 2603337. This work was performed using resources provided by the Cambridge Service for Data Driven Discovery (CSD3), operated by the University of Cambridge Research Computing Service (www.csd3.cam.ac.uk), and funded by Dell EMC, Intel, and Tier-2 funding from the Engineering and Physical Sciences Research Council (capital grant EP/T022159/1), as well as DiRAC funding from the Science and Technology Facilities Council (www.dirac.ac.uk). This work also used the DiRAC Extreme Scaling service (Tursa) at the University of Edinburgh, managed by the Edinburgh Parallel Computing Centre on behalf of the STFC DiRAC HPC Facility. The DiRAC service at Edinburgh was funded by BEIS, UKRI, and STFC capital and operations grants. DiRAC is part of the UKRI Digital Research Infrastructure. \\
\indent The simulations were run with versions \href{https://github.com/idefix-code/idefix/releases/tag/v1.1}{v1.1.0} and \href{https://github.com/idefix-code/idefix/releases/tag/v2.0.03}{v2.0.03} of \textsc{idefix}. The data were processed and plotted with \textsc{python} via various libraries, in particular \href{https://github.com/numpy/numpy}{\textsc{numpy}}, \href{https://github.com/matplotlib/matplotlib}{\textsc{matplotlib}} and \href{https://github.com/scipy/scipy}{\textsc{scipy}}, and also the following project under development: \href{https://github.com/volodia99/nonos}{\textsc{nonos}} to analyse results from \textsc{idefix} simulations.

\section*{Data availability}
The data underlying this article will be shared on reasonable request to the corresponding author.

\FloatBarrier



\bibliographystyle{mnras}
\bibliography{zotero_dec_24}







\bsp	
\label{lastpage}
\end{document}